\documentclass[twocolumn]{aastex631}

\begin{document}

\title{Out on a Limb: The Signatures of East-West Asymmetries in Transmission Spectra from General Circulation Models}

\author[0000-0002-3034-8505]{Kenneth E. Arnold}
\affiliation{Department of Astronomy, University of Maryland, 4296 Stadium Drive, College Park, MD 20742, USA}
\affiliation{Department of Astronomy, University of Wisconsin–Madison, 475 N. Charter St., Madison, WI 538706, USA}

\author[0000-0002-2454-768X]{Arjun B. Savel}
\affiliation{Department of Astronomy, University of Maryland, 4296 Stadium Drive, College Park, MD 20742, USA}

\author[0000-0002-1337-9051]{Eliza M.-R. Kempton}
\affiliation{Department of Astronomy, University of Maryland, 4296 Stadium Drive, College Park, MD 20742, USA}

\author[0000-0001-8206-2165]{Michael T. Roman}
\affiliation{Department of Astronomy, School of Physics and Astronomy, University of Leicester, Leicester, UK}

\author[0000-0003-3963-9672]{Emily Rauscher}
\affiliation{Department of Astronomy, University of Michigan, 1085 South University Avenue, Ann Arbor, MI 48109, USA}

\author[0000-0003-0217-3880]{Isaac Malsky}
\affil{Jet Propulsion Laboratory, California Institute of Technology, Pasadena, CA 91109, USA}

\author[0000-0002-6980-052X]{Hayley Beltz}
\affiliation{Department of Astronomy, University of Maryland, 4296 Stadium Drive, College Park, MD 20742, USA}

\author[0000-0001-8342-1895]{Maria E. Steinrueck}
\altaffiliation{51 Pegasi b fellow}
\affiliation{Department of Astronomy \& Astrophysics, University of Chicago, Chicago, IL, USA}

\begin{abstract}

In the era of JWST, observations of hot Jupiter atmospheres are becoming increasingly precise. As a result, the signatures of limb asymmetries due to temperature or abundance differences and the presence of aerosols can now be directly measured using transmission spectroscopy. Using a grid of general circulation models (GCMs) with varying irradiation temperatures  (1500\,K – 4000\,K) and prescriptions of cloud formation, we simulate 3D ingress/egress and morning/evening-limb transmission spectra. We aim to assess the impact that clouds, 3D temperature structure, and non-uniform distribution of gases have on the observed spectra, and how these inhomogeneities can be identified.  A second goal is to assess the relative merits of two separate methods (ingress/egress vs.\ morning/evening-limb spectroscopy) for isolating atmospheric asymmetries. From our models, it is evident that an east-west temperature difference is the leading order effect for producing ingress/egress or morning/evening-limb spectral differences. We additionally find that clouds contribute strongly to the observed limb asymmetry at moderate irradiation temperatures in our grid ($\sim2000$ K $< T_{\mathrm{irr}} < 3500$ K). At lower temperatures clouds equally dominate the optical depth on both limbs, while at higher temperatures the entire terminator region remains cloud-free.  We develop limb asymmetry metrics that can be used to assess the degree of east-west asymmetry for a given planet and predict trends in these metrics with respect to irradiation temperature that are indicative of various physical processes.  Our results are useful for predicting and diagnosing the signatures of limb asymmetries in JWST spectra.
\end{abstract}
\keywords{Exoplanet atmospheres (487); Hot Jupiters (753); Transmission spectroscopy (2133); Theoretical models (2107); Radiative transfer simulations (1967)}
\section{Introduction} \label{sec:intro}

Hot Jupiters are implicitly three-dimensional (3D) objects.  Given that they have semimajor axes $<$ 0.1 AU, these planets are highly irradiated, and they are believed to be synchronously rotating \citep[e.g.,][]{rasio1996tidal}.  The resulting strongly non-uniform heating pattern implies that they should exhibit considerable variations in temperature (and as a result, their composition, cloud properties, etc.) between their daysides and nightsides, as well as between their eastern and western hemispheres \citep[e.g.,][]{showman2002atmospheric, coopershowman05, knutson07, rauscher2010three, parmentier2016transitions, roman2019modeled}. East-west asymmetries, which are of key importance to this paper, arise because the planet's rotation and winds carry hot gas eastward from the substellar point, leading to a hotter eastern (or ``evening") limb and colder western (or ``morning") limb \citep{showman2011equatorial}.  

The impact of these 3D inhomogeneities in hot Jupiter atmospheric properties has been shown to have observable consequences on their transmission spectra \citep[e.g.,][]{fortney10_3D, flowers19, ehrenreich2020nightside,macdonald2020so, lacy20biases, zamyatina2023observability,wardenier2024phase}. To account for this non-uniformity, retrieval techniques have been developed to recover two-component (e.g., day/night or east/west) or multi-component atmospheres, enabling the recovery of distinct temperatures, abundances, and aerosols in each region \citep{lacy20biases, nixon2022aura,macdonald23poseidon}.  Other retrieval techniques have been developed to identify the presence of ``patchy" terminator cloud coverage \citep[e.g.,][]{line16patchy, macdonald2017hd,pinhas2019h2o,nixon2022aura}.  The latter approach is useful for constraining the fraction of a planet's limb that is enshrouded in aerosols. For example, analyses of the JWST transmission spectrum of the hot Jupiter WASP-39b support the presence of patchy cloud coverage and inhomogeneous temperatures in order to fully reproduce the observations \citep{feinstein2023early, chen2025asymmetry}.

However, a downside to the classical transmission spectroscopy technique is that it cannot precisely identify \textit{where} in the atmosphere any inhomogeneities arise. Another way of saying this is that transmission spectroscopy probes the entire unresolved terminator all at once (by fitting the \textit{maximum} transit depth as a function of wavelength), so any spatial information is only indirectly inferred via subtle perturbations to line profile shapes.  Swapping the properties of the eastern and western hemispheres will result in identical transmission spectra, and one cannot readily distinguish between globally patchy cloud coverage and clouds concentrated on a single hemisphere.
Recently, more direct ways of spatially ``mapping'' the terminator region have been proposed. In the era of precision transmission spectroscopy with JWST, extracting phase-resolved transmission spectra, as the planet's limb scans across the stellar disk, is an exciting new way of directly uncovering the non-uniform nature of exoplanetary atmospheres.  

Several techniques have been proposed to map the terminator through phase-resolved transmission spectroscopy. These include (i) separately extracting transmission spectra during ingress and egress phases \citep{kempton17ingressegress}, (ii) modeling the shape of the spectroscopic light curves to independently recover transmission spectra corresponding to the planet's morning and evening limbs \citep{powell2019transit, jones20catwoman, espinoza21catwoman}, and (iii) modeling the shape of the spectroscopic light curves to fully recover the effective shape of the terminator region as a function of the observed wavelength \citep{grant2023tran+}. In this paper, we will focus on and apply the concepts of (i) and (ii) from this list, which can both be most closely linked with the spectra originating from distinct atmospheric regions.  All of these transit mapping techniques are novel, and they are just now being successfully applied to data for the first time \citep{espinoza2024inhomogeneous, murphy24, tada2025probing}. We aim in this work to predict the observational signatures arising from applying these technique and also to better understand their relative strengths and shortcomings.  

To predict the signatures of limb asymmetries in transmission spectra in this work, we turn to the results of hot Jupiter general circulation models (GCMs). Such models predict possible spatial variations in temperature, chemistry, and cloudiness for a given hot Jupiter, given various underlying assumptions (e.g., chemical equilibrium, specific prescriptions for atmospheric drag, or simplifications to cloud microphysics). We focus in this work specifically on the impact of clouds in imparting observable limb asymmetries because clouds have been found to be nearly ubiquitous in all but the very hottest exoplanets \citep[e.g.,][]{charbonneau2002detection,sing2016continuum, gao2020aerosol,grant2023jwst,dyrek2024so2}. Clouds also have a first-order impact in shaping transmission spectra via the muting of spectral features and producing sloped spectra at optical wavelengths \citep[e.g.,][]{wakeford2015transmission}. The presence of clouds is expected to be highly temperature-dependent due to the nature of condensation processes, and multiple GCMs have predicted strongly spatially inhomogeneous cloud coverage in hot Jupiter atmospheres, including at the terminator \citep[e.g.,][]{parmentier2016transitions, powell2019transit, roman2019modeled}.  The radiative feedback from clouds can also significantly alter the temperature structure of a hot Jupiter's atmosphere, which may either enhance or diminish east-west temperature inhomogeneities \citep{roman2019modeled,parmentier2021cloudy, roman2021clouds, malsky2024direct}.  

In this paper, we simulate 3D transmission spectra from the grid of clear and cloudy hot Jupiter GCMs presented in \cite{roman2021clouds}. In Section \ref{sec:method}, we describe the GCMs and our techniques for calculating  simulated transmission spectra, including how we generate ingress/egress and morning/evening-limb spectra. In Section \ref{sec:results}, we present our results.  We show the terminator maps that arise from the input GCMs and their resulting spectra.  We additionally diagnose how these results depend on planetary irradiation temperature and treatment of clouds in the GCMs.  In Section \ref{sec:discussion}, we discuss the implications of our results for current and upcoming observations with JWST.  We develop a quantitative metric for determining the strength of limb asymmetries in various wavelength ranges corresponding to JWST instrument modes and important spectral features. 
Finally in Section \ref{sec:conclusion}, we conclude with a summary and some thoughts on how best to extract 3D atmospheric information from transit observations.

\begin{deluxetable*}{lccc}
\tablecolumns{4}
\tablecaption{Fixed Planetary Parameters from \cite{roman2021clouds} GCMs\label{table:parameters}}\tablehead{\colhead{Parameter} & \colhead{Value} & \colhead{Units} & \colhead{Comment}}
\startdata
Radius of the planet, R$_{p}$ & 9.65 $\times$ 10$^{7}$ & m & 1.35 R$_{Jup}$ \\
Gravitational acceleration, $g$ & 10.0 & m s$^{-2}$ & The lower-gravity models from \cite{roman2021clouds} \\
Rotation rate, $\Omega$ & 3.85 $\times$ 10$^{-5}$ & s$^{-1}$ & Spin-synchronized, 1.89 day orbit\\
& & &\\
\enddata
\end{deluxetable*}

\section{Methods} \label{sec:method}
Our modeling framework begins with the \citet{roman2021clouds} grid of GCMs, which we post-process with a ray-striking radiative transfer code \citep{kempton2012constraining} to produce ingress/egress spectra and morning/evening spectra.  Details are described, below.

\subsection{3D Hot Jupiter GCMs\label{subsec:GCMs}}
The \cite{roman2021clouds} models (which used the RM-GCM code; \citealt{rauscher2012general}) solve the primitive equations of meteorology and are coupled to a double-gray, two stream radiative transfer scheme based on \cite{toon1989rapid}. One important aspect of the \cite{roman2021clouds} GCMs that has also been considered in other works \citep{lee2016dynamic, lines2018simulating, lines2019overcast} is that they include radiative feedback from cloud layers that are allowed to form in the atmosphere in chemical equilibrium over a wide range of irradiation temperatures, using the formalism first developed in \citet{roman2017modeling, roman2019modeled}. Briefly, the GCMs include the scattering and absorption from 13 condensible species listed in Table 3 of \cite{roman2021clouds}. The full GCM grid explores sensitivity of assumptions related to surface gravity, condensation efficiency, and cloud thickness for a range of irradiation temperatures ($1500 - 4000$ K). In this paper we only focus on a subset of GCMs as we are just concerned with the parameters primarily affecting the transmission spectra i.e., temperature and cloud treatment. The relevant planetary properties from the \cite{roman2021clouds} GCMs are replicated in Table \ref{table:parameters}, we only consider models with a surface gravity of \mbox{g = $10$ m s$^{-2}$} (as lower-gravity planets are more amenable to transmission spectroscopy; e.g., \citealt{kempton17ingressegress}) and a condensible mass fraction of $f = 10$\%. 

\begin{figure}
    \centering
    \includegraphics[scale=0.5]{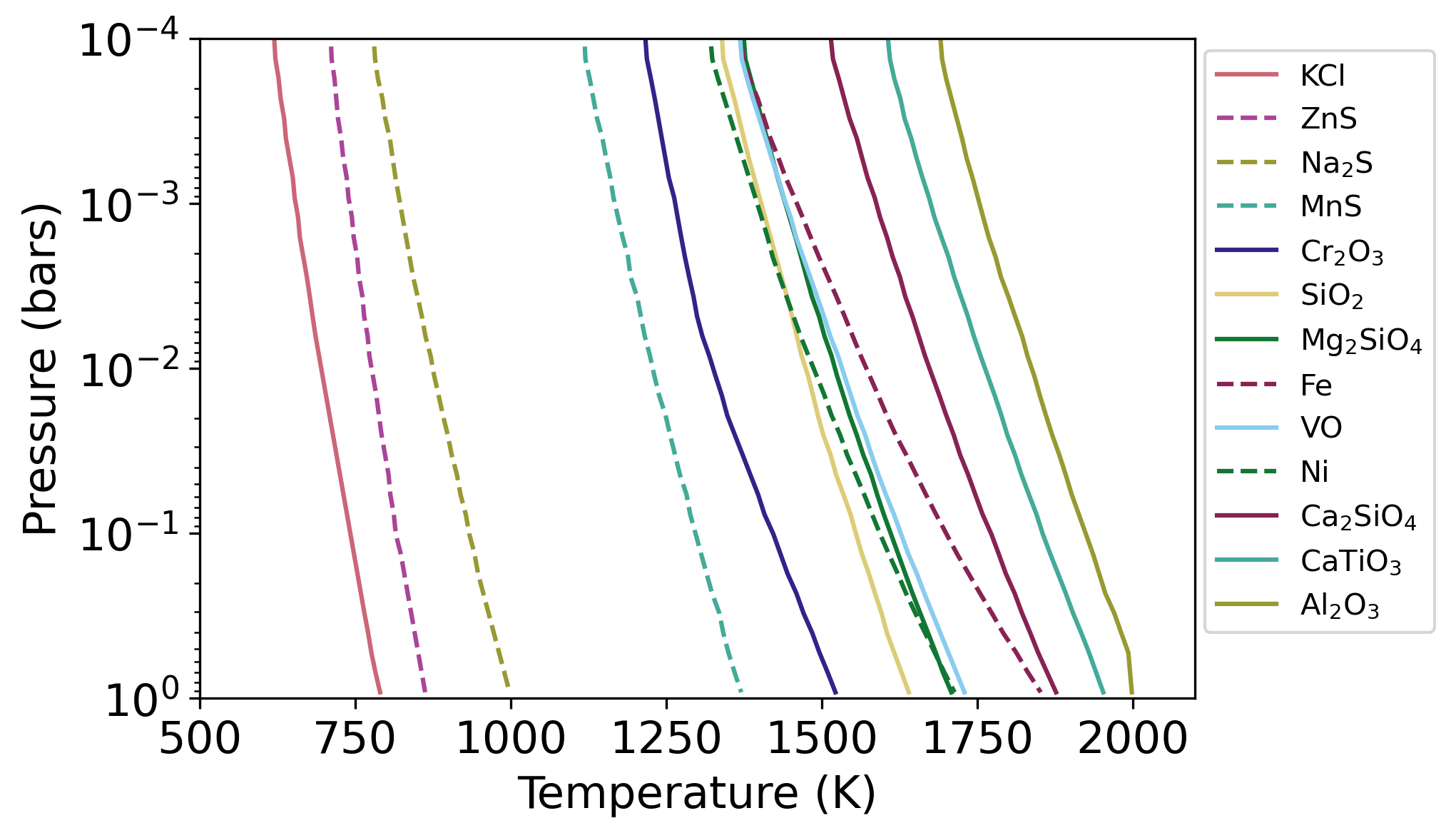}
    \caption{Condensation curves of the 13 condensable species calculated from \cite{mbarek2016clouds} with nucleation-limited species plotted with dashed lines.}
    \label{fig:condensation curves}
\end{figure}

\begin{figure*}
    \centering
    \includegraphics[scale=0.45]{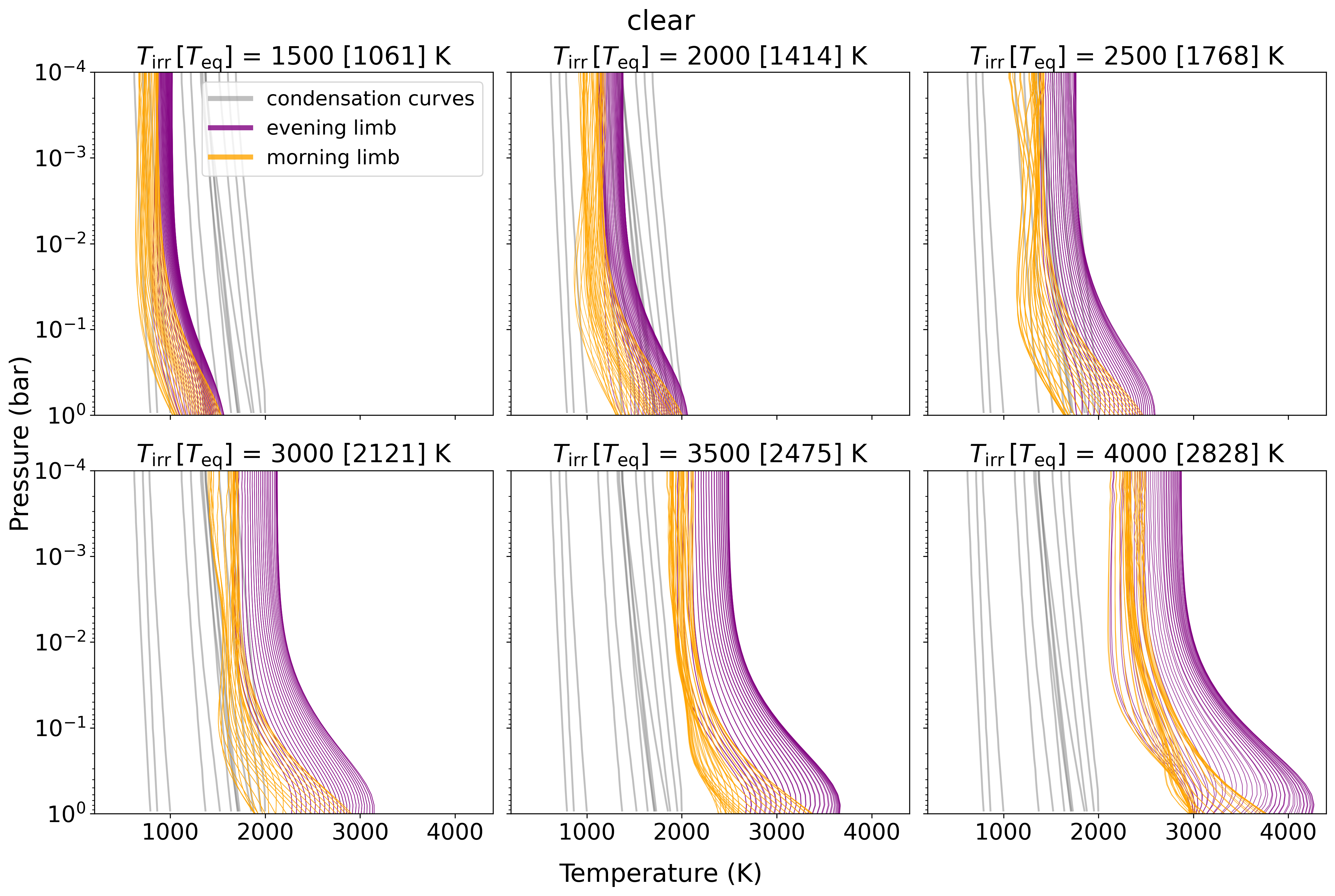}
    \caption{Vertical temperature--pressure profiles of the clear GCMs at 6 different irradiation temperatures. In gold are the TP profiles for the morning (western) terminator, at the center of ingress, for a single longitude of 270$\degr$. In purple are the TP profiles for the evening (eastern) terminator, at the center of egress, for a single longitude of 90$\degr$. In gray we plot the condensation curves of the 13 condensable species from \cite{roman2021clouds}. Because these are the clear-atmosphere models, there is no condensation; thus we plot condensation curves in gray to demonstrate where we would expect clouds to form as a comparison to the cloudy models. With increasing irradiation temperature, there is a wider range of temperatures for the evening limb, while the morning limb consistently shows less variation in temperature.}
    \label{fig:clear TP-profile}
\end{figure*}

\begin{figure*}
    \centering
    \includegraphics[scale=0.45]{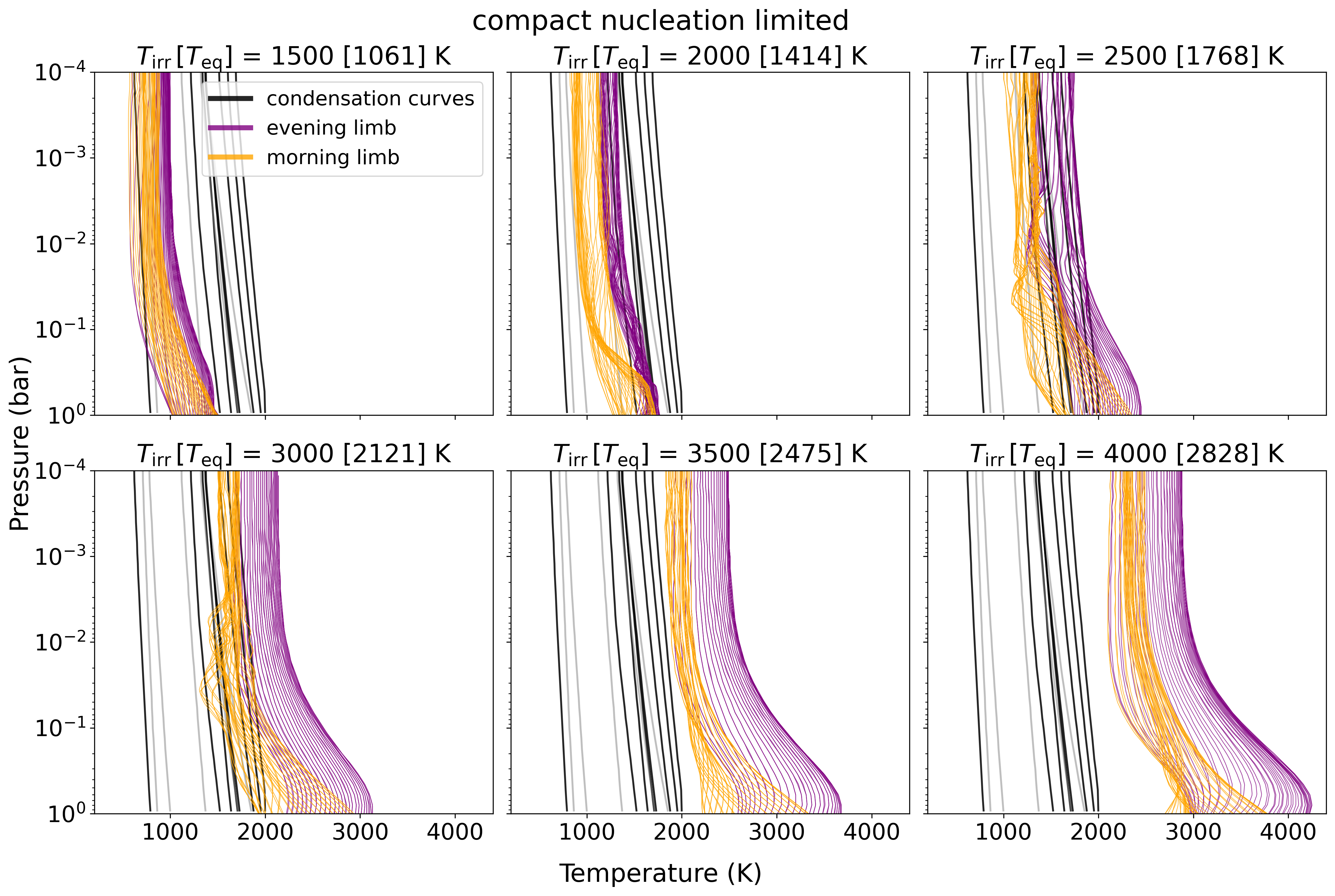}
    \caption{Same as Figure~\ref{fig:clear TP-profile} but for the compact nucleation-limited GCMs. Condensation curves for clouds we include in the compact nucleation limited GCMs are plotted in black, while condensation curves for clouds we don't include are plotted in gray. The same is true for Figures \ref{fig:extended TP-profile}, \ref{fig:compact TP-profile}, \ref{fig:nuclim TP-profile}. For the highest irradiation temperatures, the TP profiles are all too hot for condensation to occur on either limb.  Moving to lower irradiation temperatures, condensation first occurs on the morning limb, followed by the evening limb.  Radiative feedback from clouds is evident in the ``wobbling'' of the TP profiles and by looking for departures of these TP profiles from the ones shown in Figure~\ref{fig:clear TP-profile}. In these models, the clouds impact the  most strongly at $T_{\mathrm{irr}} = 2500\textnormal{K}$.}
    \label{fig:comnuclim TP-profile}
\end{figure*}

The clouds in the \citet{roman2021clouds} GCMs are implemented with various simplifying assumptions that are aimed at approximating the outcomes of more computationally-intense cloud microphysics models.  The clouds form in chemical equilibrium wherever the local temperature (at a given pressure) drops below the condensation temperature for a given cloud species, using the condensation curves for a solar composition atmosphere from \citet{mbarek2016clouds}. Moreover, the condensates are modeled as Mie scatterers with a vertical gradient in particle size \citep{de1984expansion, mishchenko1999bidirectional}; the cloud particle size increases exponentially with increasing pressure, introducing a pressure and wavelength dependence in the clouds' extinction efficiencies. Figure \ref{fig:condensation curves} shows the condensation curves for each of the condensable species included in our cloudy GCMs.

\begin{deluxetable*}{lccccc}
\tablecolumns{6}
\tablecaption{Orbital and stellar parameters for ingress / egress spectra  \label{table:orbit}} \tablehead{\colhead{T$_{\mathrm{irr}}$ (K)} & \colhead{stellar radius ($R_{\odot}$)} & \colhead{semi-major axis (AU)} & \colhead{T$_{I}$, T$_{IV} (\degr$)} & \colhead{T$_{II}$, T$_{III} (\degr$)}} 
\startdata
1500 & $0.52$ & $6.91 \times 10^{-2}$ & $\pm4.91$ & $\pm2.83$  \\ 
1750 & $0.66$ & $5.07 \times 10^{-2}$ & $\pm6.37$ & $\pm4.15$  \\ 
2000 & $0.81$ & $3.88 \times 10^{-2}$ & $\pm8.05$ & $\pm5.70$  \\ 
2250 & $0.98$ & $3.07 \times 10^{-2}$ & $\pm9.96$ & $\pm7.47$  \\ 
2500 & $1.16$ & $2.49 \times 10^{-2}$ & $\pm12.1$ & $\pm9.49$  \\
2750 & $1.35$ & $2.05 \times 10^{-2}$ & $\pm14.5$ & $\pm11.7$  \\ 
3000 & $1.54$ & $1.73 \times 10^{-2}$ & $\pm17.1$ & $\pm14.2$  \\ 
3250 & $1.75$ & $1.47 \times 10^{-2}$ & $\pm20.0$ & $\pm16.9$  \\
3500 & $1.97$ & $1.27 \times 10^{-2}$ & $\pm23.1$ & $\pm19.9$  \\ 
3750 & $2.20$ & $1.10 \times 10^{-2}$ & $\pm25.6$ & $\pm23.2$  \\ 
4000 & $2.43$ & $9.71 \times 10^{-3}$ & $\pm30.2$ & $\pm26.8$  
\enddata
\tablenotetext{}{These parameters were chosen so that the planet radius $R_{P} = 1.35$ $R_{Jup}$ and orbital period $P = 1.89$ days remain constant for varying $T_{\mathrm{irr}}$.  The last two columns describe orbital phases of points of contact (i.e., beginning of ingress, end of egress, end of ingress, and beginning of egress).  Spectra are simulated at 5 equally-spaced phases during both ingress and egress. }
\end{deluxetable*}

In the absence of a full microphysics model, the \citet{roman2021clouds} GCM grid explores five different cloud treatments in their work.  The first is a fully cloud-free case, in which no clouds are allowed to form. For the cloudy cases, the GCMs then explore different end-member scenarios for cloud thickness. The vertical extent of clouds can depend on vertical mixing strength, with stronger mixing allowing thicker clouds; given that vertical mixing in hot Jupiter atmospheres is unknown to orders of magnitude \citep[e.g.,][]{moses2013compositional,parmentier20133d,agundez2014pseudo}, the two end-member cloud cases correspond to stronger and weaker mixing. The first is an ``extended'' cloud model, in which the clouds are allowed to vertically extend all the way to the topmost layer of the atmosphere in any location (defined by latitude and longitude) above where clouds are able to first form at lower altitudes, so long as the atmosphere is still cold enough.  The second is a ``compact'' cloud model, in which clouds extend vertically by exactly one gas scale height above the location where they form. Microphysical considerations likely limit cloud height \citep[e.g.,][]{powell2019transit}, so the compact cloud model case is likely more physically realistic. 

The GCM grid furthermore spans two scenarios for cloud \textit{composition}, with the aim of exploring the wide range of possible condensates in exoplanet atmospheres \citep[e.g.,][]{mbarek2016clouds}.  In keeping with \cite{roman2021clouds}, we consider the baseline case to be the one in which all 13 condensible species are allowed to form clouds.  The second case is one in which several cloud species are removed from the calculation entirely, based on their expected low particle nucleation rates \citep{gao2020aerosol}. The clouds that are removed in this ``nucleation-limited'' scenario are ZnS, Na$_2$S, MnS, Fe, and Ni. Furthermore, the nucleation-limited clouds are simulated for both the extended and compact cloud thicknesses. In summary, there are five different cloud treatments explored by the \citet{roman2021clouds} GCMs, which are termed clear, extended, compact, extended nucleation-limited, and compact nucleation-limited. 

Figures \ref{fig:clear TP-profile}, \ref{fig:comnuclim TP-profile}, \ref{fig:extended TP-profile}, \ref{fig:compact TP-profile}, and \ref{fig:nuclim TP-profile} show the temperature-pressure (TP) profiles of the morning and evening limbs  (approximately corresponding to ingress and egress respectively) for each of the five cloud treatments. In our subsequent modeling, we focus most strongly on the compact nucleation-limited clouds because they are deemed to be the most physically realistic scenario. This is based on the degree to which they are able to reproduce existing observations and match the outputs of cloud microphysics models \citep[see further discussion in][]{powell2019transit, gao2020aerosol, roman2021clouds}.  The other cloud treatments are included as points of comparison.

To prepare the GCMs for post-processing with our radiative transfer code (Section \ref{subsec:rt}), we interpolate them to a constant altitude grid. We must do this because the radiative transfer calculation requires striking straight-line rays through the atmosphere along the observer's sight line.  Because different regions of the planetary atmosphere have different scale heights, ray-striking through the GCMs'  native constant pressure grid would not follow straight lines through physical space. When mapping from pressure to altitude, we truncate the atmosphere at the top of the GCM domain (approximately an isobar of $5.7 \times 10^{-5}$ bar).  This results in the hotter regions of the atmosphere extending to higher altitudes than the cooler regions.  Numerically, this step is accomplished by setting the temperature and gas opacity to zero in our fixed altitude grid for regions outside of the GCM model domain. 

Our mapping of pressure onto altitude makes three assumptions that are consistent with the GCMs:

\begin{enumerate}
    \item  hydrostatic equilibrium. This is an appropriate assumption, as simulations of outflows indicate that they are launched at lower pressures than those spanned by our models \citep[e.g.,][]{murray2009atmospheric}.
    \item constant gravity. Gravity changes by less than 10\% in the regions of the atmospheres probed in transmission for our very hottest models, and substantially less for the majority of our grid.
    \item constant mean molecular mass. The hottest exoplanets exhibit significant hydrogen dissociation and recombination throughout the atmosphere \citep{bell2018increased, tan2019atmospheric}, resulting in strong variations in mean molecular mass through their atmospheres. \texttt{FastChem} \citep{stock2018fastchem} calculations indicate that ignoring this effect negligibly affects our cooler models. While our hottest models reach temperatures at which $\rm H_2$ dissociation becomes non-negligible, this physics was not implemented in the \cite{roman2021clouds} GCMs (along with other effects specific to high-temperature atmospheres). We therefore omit mean molecular mass variations from our post-processing to maintain self-consistency between our models and the GCMs on which they are based.
\end{enumerate}

Furthermore, we choose a single iso-altitude contour as the base of our atmosphere. This iso-altitude corresponds to the deepest altitude (smallest planetary radius) corresponding to an atmospheric pressure of 1 bar, along any of the individual 1-D T-P profiles.  This means that the atmosphere extends to a pressure of at least 1 bar at every location. We remove regions of the atmosphere with pressures greater than this value to save on computation time because the atmosphere is fully optically thick in transmission geometry in these locations, meaning that they do not contribute any relevant information to the spectrum.

\subsection{Transmission Spectra Radiative Transfer\label{subsec:rt}}
We simulate transmission spectra at a resolution of $R\sim300$ over the wavelength range of $0.5- 10$ $\mu$m, roughly corresponding to JWST capabilities. The ray-striking radiative transfer code we employ is the one from \citet{kempton2012constraining} and \citet{flowers19}, which was more recently updated in \citet{savel2022no}.  In this work, we add the capability to account for the wavelength dependence of the cloud optical properties for a set of $N$ individual cloud species ($N = 13$ or 8 for the baseline and nucleation-limited clouds in this study, respectively), based on their local abundances in the atmosphere.


Our approach to implementing wavelength-dependent cloud extinction is similar to \cite{harada2021signatures}, who post-processed the GCMs from \citet{roman2019modeled} but for high-resolution ($R \sim 10^{6}$) emission spectra.
Our general approach is to calculate transmission spectra from the GCM output by summing the intensity of incident rays from the background star that pass through the planetary atmosphere. The emergent intensity is calculated using the extinction-only radiative transfer equation for transmission
\begin{equation}
\label{eq:transmit}
    I_{\lambda} = I_{0_{\lambda}} e^{-\tau_{\lambda}},
\end{equation}
where $I_{0_{\lambda}}$ is the incident intensity from the star entering the planetary atmosphere and $\tau_{\lambda}$ is the line-of-sight optical depth of the atmosphere. We treat scattering (e.g., Rayleigh scattering) as pure extinction, and we ignore any effects of additional scattering into the beam or gas refraction, both of which have been found to be of minimal effect to hot Jupiter transmission spectroscopy \citep{hubbard2001theory,hui2002atmospheric,betremieux2015refraction,robinson2017analytic}.

The line-of-sight optical depth is defined as 
\begin{equation}
\label{eq:optical}
    \tau_\lambda = \int(\kappa_{\mathrm{gas,\lambda}} + \kappa_{\mathrm{cloud,\lambda}}) dl,
\end{equation}
where $\kappa_{\mathrm{gas,\lambda}}$ is the wavelength-dependent gas opacity at local atmospheric temperature and pressure, $\kappa_{\mathrm{cloud, \lambda}}$ is the wavelength-dependent cloud opacity, and $dl$ is the line-of-sight path length through each atmospheric grid cell encountered by a light ray. The gas opacity is a product of the gas cross-section and its number density, summed up over all contributing gas species. We calculate abundances at solar metallicity and C/O \cite{lodders2003solar} with \texttt{FastChem} \citep{stock2018fastchem}. We do not consider depletion of gas abundances by cloud formation.  This should have negligible impact on the key molecular absorbers in the IR (i.e. H$_{2}$O, CO, CO$_{2}$, CH$_{4}$), since the atmosphere is never cold enough for any of these species to condense. For gas cross-section, we include C$_{2}$H$_{2}$ \citep{chubb2020exomol}, CH$_{4}$ \citep{yurchenko2014exomol}, CO$_{2}$ \citep{yurchenko2020exomol}, CO \citep{li2015rovibrational}, H$_{2}$O \citep{polyansky2018exomol}, H$_{2}$S \citep{azzam2016exomol}, HCN \citep{barber2014exomol}, K \citep{kurucz2011including}, Na \citep{kurucz2011including}, NH$_{3}$ \citep{coles2019exomol}, FeH, TiO \citep{mckemmish2019exomol}, and VO \citep{mckemmish2016exomol}.

To calculate cloud absorption, we begin with the expression for \textit{radial} cloud optical depth from \cite{roman2019modeled}:

\begin{equation}
    \mathrm{d} \tau_{\rm cloud,\,\lambda,\,radial} = \frac{3m_gQ_{e, \lambda}f}{4r\rho},
\end{equation}
where $m_g$ is the cloud mass in a layer, $Q_{e, \lambda}$ is the particle-scattering extinction efficiency, $r$ is the cloud particle radius (set by the pressure-dependent particle radius profiles of \citealt{roman2021clouds}), and $\rho$ is the cloud particle density.\footnote{The GCM technically only tracks the cloud optical depth at one wavelength, as it is a double-gray simulation. So, to calculate the wavelength-dependent optical depth, we rescale the ``gray'' optical depth by the ratio of the ``gray'' $Q_{e}$ and the non-gray $Q_{e, \lambda}$.} This optical depth is tracked in the GCM to compute cloud temperature feedback. Densities, visible refractive indices, and thermal indices used to calculate the particle-scattering extinction efficiencies are listed in Table 3 from \cite{roman2021clouds}. We then convert this GCM-tracked quantity to $\kappa_{\mathrm{cloud, \lambda}}$ for our post-processing:

\begin{equation}
     \kappa_{\rm cloud,\,\lambda} = \frac{\mathrm{d}\tau_{\rm cloud,\,\lambda,\,radial}}{\mathrm{d}s},
\end{equation}
for layer thickness $\mathrm{d}s$. Values for $\kappa_{\rm cloud,\,\lambda}$ are calculated for each of the 13 cloud species we consider in the GCMs, and the total cloud opacity is taken as the sum of each cloud species' opacity.


In order to generate phase-dependent transmission spectra, we follow the process outlined in \cite{savel2022no} to both rotate the planet and ``back-light'' the planet and its atmosphere with an appropriately limb-darkened star. The latter is accomplished by rotating the GCM's longitude grid at each modeled orbital phase, assuming tidal synchronization. For the optically thick core of the planet (out to the base-of-atmosphere radius $R_p$), we generate a ``no-atmosphere'' transit light curve using the \texttt{batman} package \citep{kreidberg15_batman}.  We employ the quadratic limb darkening factors (LDFs) for WASP-76b from \cite{ehrenreich2020nightside} under the simplifying assumption that they do not vary with wavelength. For the atmospheric transmission component of our ingress/egress spectra, we identify the region of the limb-darkened star from which each straight-line ray originates, and apply the appropriate LDF to its incident intensity, $I_0$, prior to allowing the ray to pass through the atmosphere.  For regions of the terminator that are not in front of the star during a particular ingress/egress phase, the incident intensity is set to $I_0 = 0$. Finally, for the morning/evening limb spectra, we simply block off one half of the planet's atmosphere, depending on which limb we are simulating. Thus, no LDF is used in the calculations for the morning/evening limb spectra.


We generate transmission spectra from the grid of hot Jupiter GCMs for all 11 irradiation temperatures presented in \cite{roman2021clouds}. For the ingress/egress spectra, we calculate transmission spectra in the planet's rest frame at 5 evenly spaced orbital phases between the first and second contact of transit (ingress) and 5 orbital phases between third and fourth contact of transit (egress), as shown in Table \ref{table:orbit}. The first and last points of contact are not included in our calculations as no part of the planet would be back-lit at these orbital phases. Instead, we start with an orbital phase slightly after first contact and an orbital phase slightly before fourth contact. We then take the average of all 5 spectra as the resulting ingress/egress spectra. We take the eccentricity and impact parameter to be zero for simplicity (as also assumed in the GCMs). The orbital phases for ingress and egress are determined using Kepler's third law, holding the period of the orbit and the radius of the planet constant while varying the radius of the star and the semi-major axis. This approach is required to retain self-consistency with the irradiation temperature and rotation rate from the GCM grid, as both of these variables strongly control hot Jupiter circulation \citep{showman2002atmospheric,menou2003weather,cho2008atmospheric,showman2009atmospheric,rauscher2014atmospheric,showman2015three, komacek2016atmospheric,penn2017thermal,mendoncca2020angular,roth2024hot}.  Note that this process results in the hottest planets rotating through a much larger range of orbital phase angles during transit ($\sim60^\circ$) compared with the coldest planets ($\sim10^\circ$), as is expected to be the case for typical transiting systems.

\begin{figure*}
    \centering
    \includegraphics[width=1.0\textwidth]{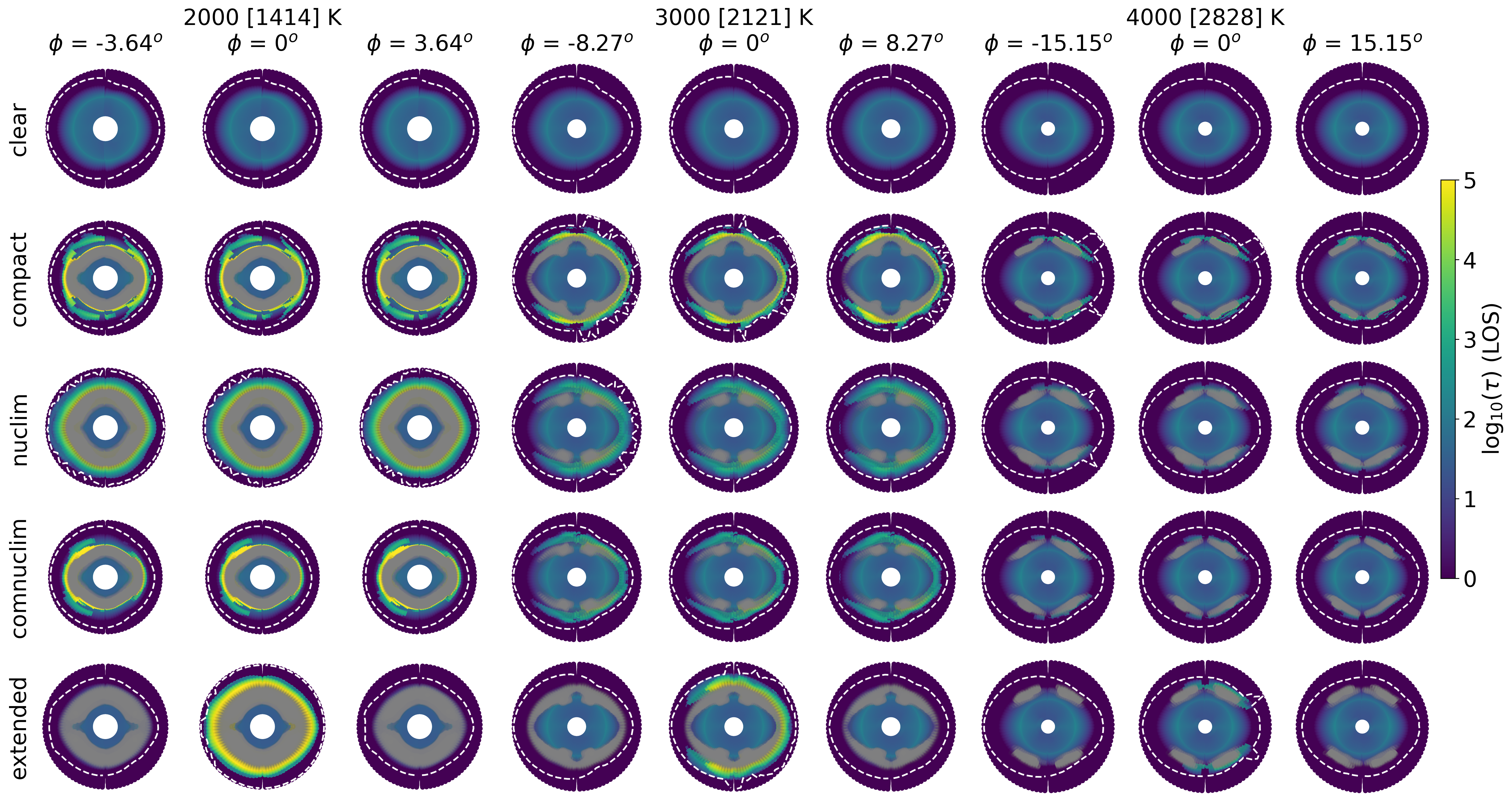}
    \caption{Line-of-sight optical depth maps for $T_{\mathrm{irr}}[T_{\mathrm{eq}}] = 2000[1414]\textnormal{K}, 3000[2121]\textnormal{K}, 4000[2828]\textnormal{K}$, at a single wavelength of 1.4 $\mu$m (water band), for each cloud treatment we consider. In this transit orientation the left side of the planet is the evening terminator and the right the morning terminator. The color scale shows the optical depth from gas, while the opacity from clouds is in a transparent gray scale (i.e., a cloud optical depth of zero would be entirely transparent and at max cloud optical depth it would be entirely gray). For each irradiation temperature, we show 3 orbital phases: one for the center of ingress, one for the center of transit, and one for the center of egress. In addition to optical depth, we also plot the height at which the optical depth reaches $\approx \frac{2}{3}$, indicating the approximate 1.4-$\mu$m photosphere. For the lower-temperature cloudy models, the optical depth is mainly dominated by the cloud opacity, while the high-temperatures models are dominated by gas opacity. Note that these plots are not to scale. The planet's atmosphere is stretched in the radial direction to make viewing easier for the reader. }
    \label{fig:optical depth grid}
\end{figure*}

\begin{figure*}
    \centering
    \includegraphics[scale=0.6]{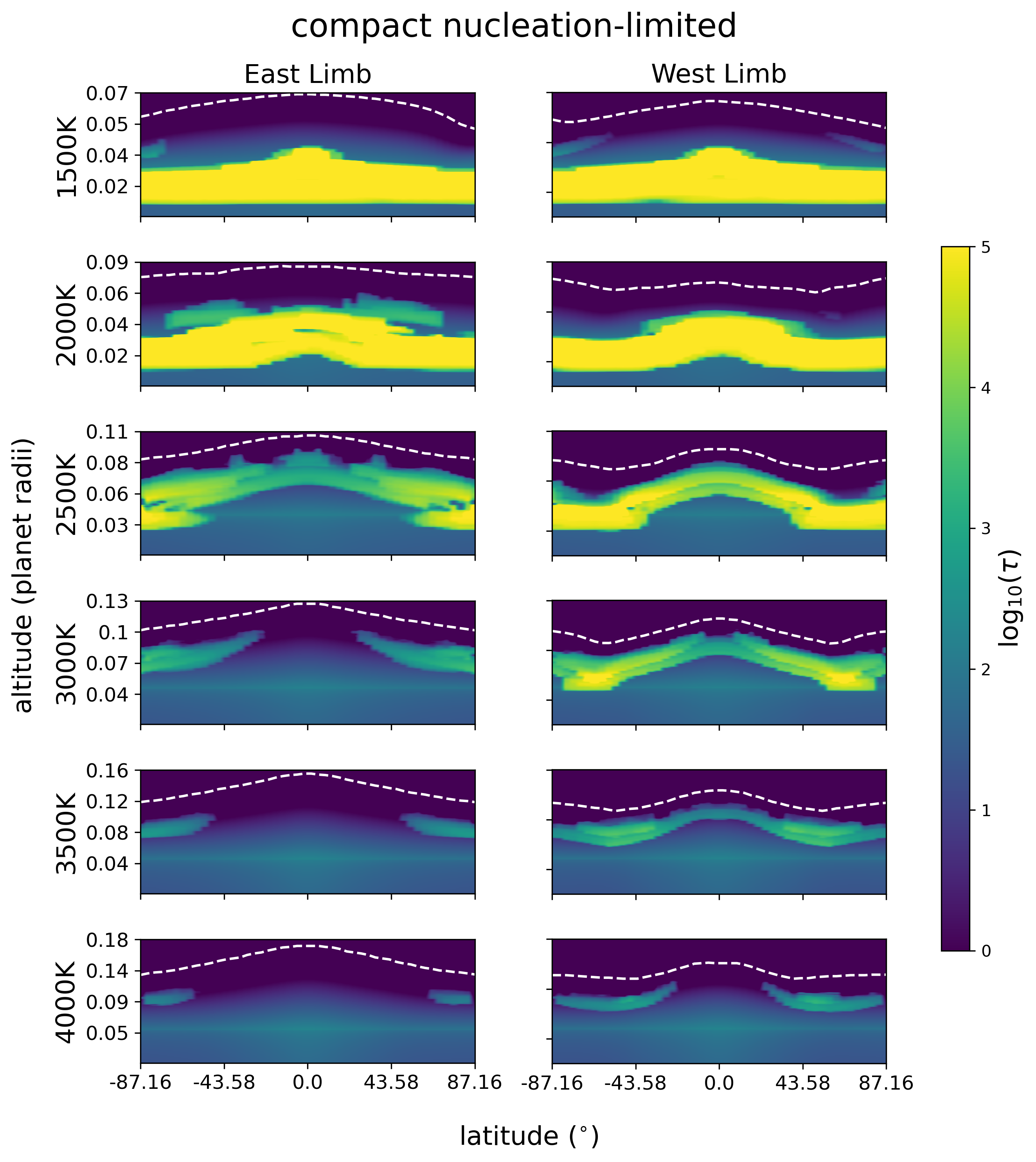}
    \caption{Optical depth maps in cartesian coordinates for $T_{\mathrm{irr}} = 1500\textnormal{K} - 4000\textnormal{K}$, at a single wavelength of 1.4 $\mu$m (water band) for the compact nucleation-limited cloud models. In this figure, the color scale indicates the combined line-of-sight optical depth from gas and clouds. For each irradiation temperature, the east (evening) limb and west (morning) limb are depicted at the center of transit. The white dashed line is the $\tau = 2/3$ surface. For the lowest irradiation temperatures, we see a very homogeneous distribution of cloud optical depth on both limbs. Regions of very high optical depth ($\gtrsim 10^4$, corresponding to bright yellow regions on the plot) are typically due to the presence of thick cloud layers.}
    \label{fig:ODM cartesian comnuclim}
\end{figure*}

\begin{figure*}
    \centering
    \includegraphics[scale=0.345]{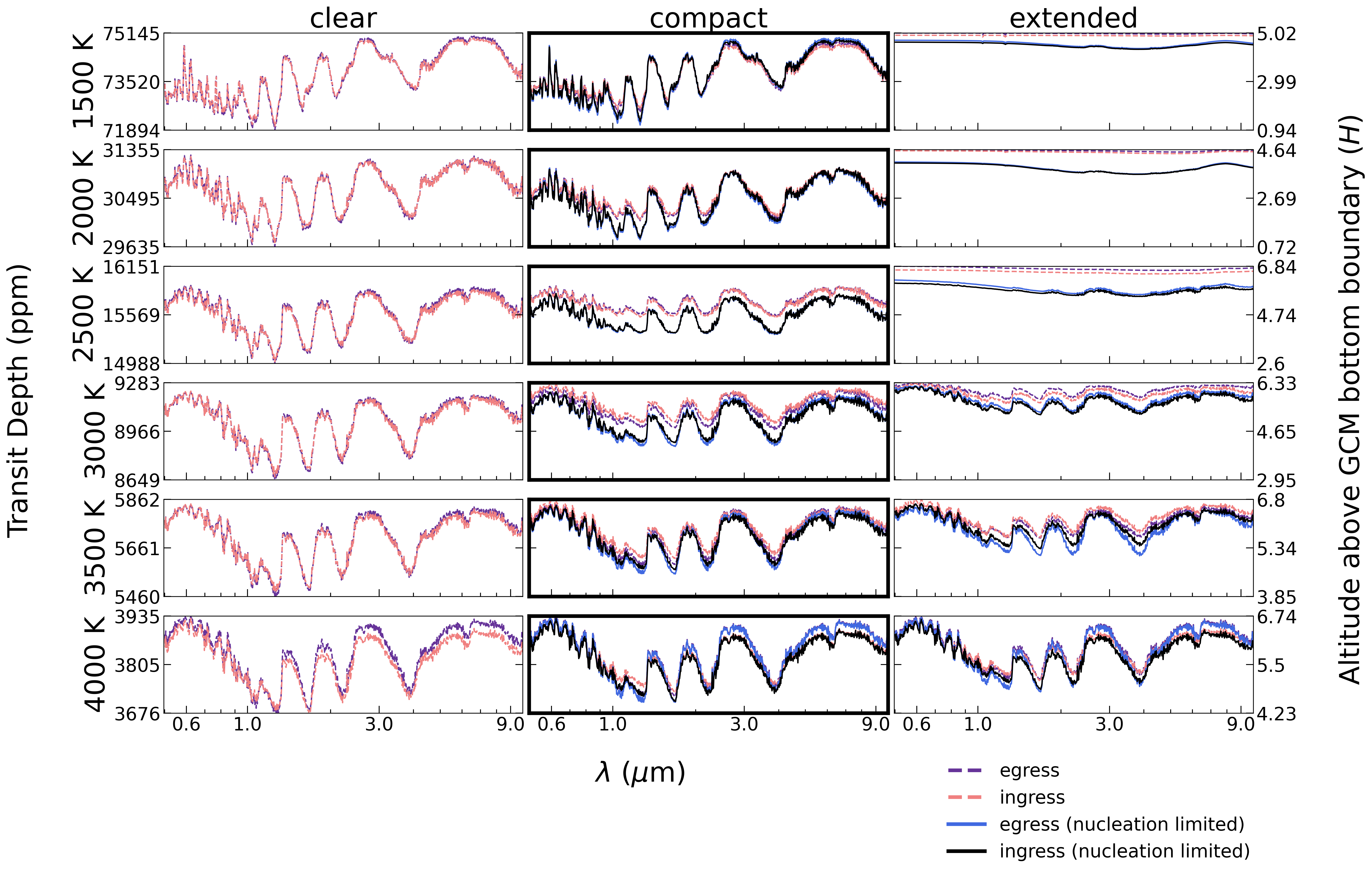}
    \caption{Ingress (magenta) and egress (orange) spectra for GCMs with $T_{\mathrm{irr}} = 1500\textnormal{K} - 4000\textnormal{K}$ for each cloud treatment we consider. The nucleation-limited cloud models have ingress and egress plotted in black and purple solid lines, respectively. Each spectrum plotted is the average of transmission spectra calculated at 5 equally-spaced ingress (or egress) phases, as indicated in Table~\ref{table:orbit}.  The ingress (egress) spectra primarily probe the planet's morning (evening) limb.  However, due to the transit geometry, there is some pollution by the opposing limb in each case, as both limbs are occulting the star during contacts 1.5--2 and 3--3.5.  Descriptions of the different cloud treatments can be found in Section~\ref{subsec:GCMs}.  The compact cloud models are highlighted because these are the most physically-motivated among the models we present. Y-axes represent scales in both transit depth in ppm (left axis) and atmospheric scale heights above the 1-bar radius, calculated with the planetary irradiation temperature} (right axis).
    \label{fig:transmission grid}
\end{figure*}

\section{Results} \label{sec:results}
We now assess the degree of limb asymmetry in the hot Jupiter GCMs as a function of temperature and cloudiness.
In Sections~\ref{sec:Temperature dependence} and \ref{sec:dependence on cloud treatment} we focus primarily on our simulated ingress/egress spectra, first examining the temperature dependence of limb asymmetry, and then the impact of various cloud treatments. In Section~\ref{sec:limb only} we discuss the results of our alternate method of extracting 3D morning/evening limb spectra. 

\subsection{Temperature dependence}\label{sec:Temperature dependence}

Optical depth maps for a subset of the GCMs are shown in Figure \ref{fig:optical depth grid} as a physical projection of the terminator region, and in Figures \ref{fig:ODM cartesian comnuclim}, \ref{fig:ODM cartesian extended}, \ref{fig:ODM cartesian compact}, and \ref{fig:ODM cartesian nuclim} in Cartesian coordinates of terminator altitude vs.\ latitude. In all cases, we track the $\tau = 2/3$ surface, indicating the approximate location of the photosphere at the wavelength of $\lambda = 1.4$ $\mu$m. This wavelength lies in the center of a strong water absorption band, so adjacent wavelengths with less opacity will produce a deeper (i.e., higher pressure) photosphere under clear atmosphere conditions.  For cloudy atmospheres, the photosphere occurs higher up (i.e., lower pressure), if the optically thick cloud top extends above the location of the clear atmosphere $\tau = 2/3$ surface. In Figure \ref{fig:optical depth grid}, we plot the cloud and gas opacities separately to demonstrate cloud coverage as a function of temperature and location along the terminator. 

There are two main temperature-dependent effects that both impact limb asymmetry.  The first is the temperature asymmetry that results in the eastern (evening) limb being hotter than the western (morning) limb.  This comes about due to the strong dayside radiative forcing of these tidally-locked planets and the subsequent eastward heat transport by the strong equatorial jet.  The temperature asymmetry can be seen in Figures~\ref{fig:optical depth grid} and \ref{fig:ODM cartesian comnuclim} by the higher photosphere on the ``puffier'' eastern limb, due to the larger pressure scale height for the hotter gas.  This is especially apparent in the clear atmosphere models, in which the cloud opacity does not contribute to the overall optical depth. 

To first order, the location and existence of clouds is also set by the local temperature of the gas.  Below an irradiation temperature of $\sim$3000 K, clouds first appear around the poles and on the cooler western limb, before ultimately covering the entire terminator region (in addition to the nightside), as a function of decreasing planetary irradiation.  This behavior is generally seen across all of the cloud models, but we will discuss key differences between the different cloud implementations in Section~\ref{sec:dependence on cloud treatment}.  The onset of clouds vs.\ irradiation temperature can be seen in Figures~\ref{fig:comnuclim TP-profile} -- \ref{fig:ODM cartesian comnuclim} and \ref{fig:extended TP-profile} - \ref{fig:ODM cartesian nuclim}.  A second-order effect arises due to the radiative feedback from the clouds, which alters the circulation and thermal structure of the atmosphere creating differences at the terminator that can be seen in Figures~\ref{fig:comnuclim TP-profile} and \ref{fig:extended TP-profile} - \ref{fig:nuclim TP-profile}. This can lead to either slight enhancement or reduction of the east-west temperature gradient and can also raise or lower the absolute temperature of the gas across the terminator region.  We generally find though that radiative feedback from clouds is less important than irradiation (which drives the east-west temperature gradient in the gas) and degree of cloudiness in setting the overall amount of limb asymmetry.  

The rotation of the planet also plays a role in generating a perceived limb asymmetry. Early in the transit (i.e., during ingress) more of the nightside of the planet will be in view, whereas later in transit the hot dayside has begun to rotate into view. More irradiated planets that orbit closer in to their host stars will rotate through a greater range of orbital phase during transit (see Table~\ref{table:orbit}), and therefore the rotation effect will be more apparent for such objects \citep{wardenier2022all}. This effect is mostly due to a hot spot offset. Proportionally, the nightside makes up less of the egress spectrum if there is an eastward hot spot offset, which is expected for hot Jupiters \citep{penn2017thermal}. This effect can be seen in Figure~\ref{fig:optical depth grid}, in which the optical depth maps for ingress vs.\ egress phases are most distinct at the highest irradiation temperatures.  

\begin{figure*}
    \centering
    \includegraphics[scale=0.35]{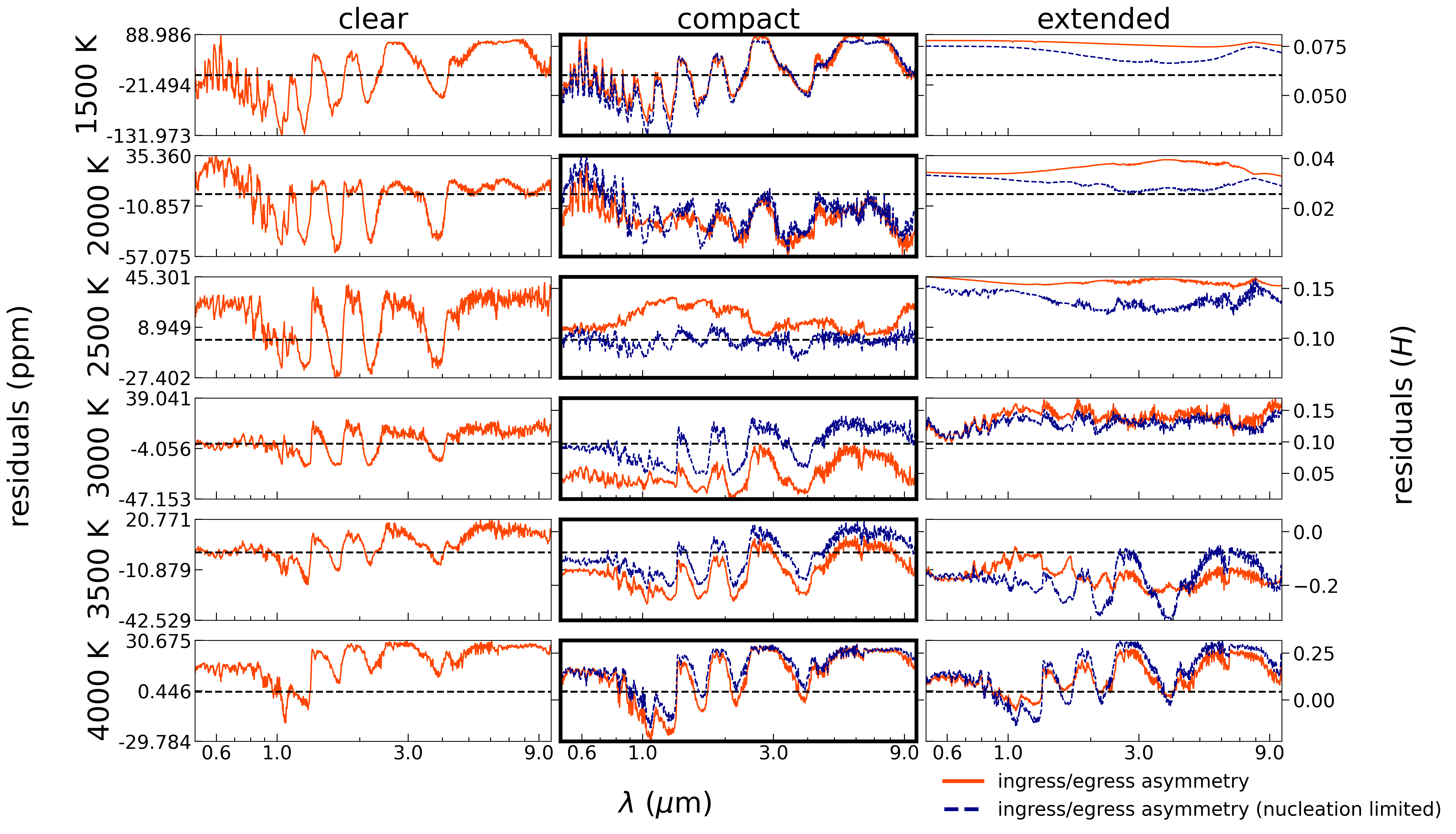}
    \caption{Difference in transit depth between egress and ingress for the spectra shown in Figure~\ref{fig:transmission grid}. Plotted in orange are the residuals for the clear and all-cloud models.  Plotted in dark blue are the residuals for the nucleation-limited models. The dashed black line is plotted to indicate zero. Positive values indicate that the egress transit depth is larger than the ingress depth (and vice versa).}
    \label{fig:residuals grid}
\end{figure*}

Figure~\ref{fig:transmission grid} shows the ingress/egress transmission spectra that result from our models, and Figure~\ref{fig:residuals grid} shows the difference between ingress and egress spectra for each case. From Figure~\ref{fig:transmission grid}, the (relative) difference between ingress and egress spectra is most pronounced for the moderately irradiated planets, where the east-west temperature difference is greatest. The rotation of the planet during transit further enhances the degree of asymmetry seen between ingress and egress phases.  From Figure~\ref{fig:residuals grid}, we see that, at most wavelengths, the egress transit depth is typically larger than that of ingress.  This is expected, as the hotter (and therefore larger scale height) gas generally lies on the evening limb, which is the side of the planet primarily probed at egress.

\subsection{Dependence on clouds}\label{sec:dependence on cloud treatment}
In Section \ref{sec:Temperature dependence}, we stated that the limb asymmetry is primarily governed by temperature, as the difference in scale height and cloud distribution between the morning and evening limb is strongly determined by the irradiation temperature of the GCMs. Here we discuss how the clouds themselves impact the degree of limb asymmetry.  In Figure~\ref{fig:transmission grid}, we can see the impacts of clouds on the ingress/egress transmission spectra become increasingly apparent in models below $T_{irr} = 3500$ K, as cloud species condense and cover the planetary limb.    

There are noticeable differences between models with extended vs.\ compact clouds. By definition, the extended clouds extend to the very top of the model domain.  Once they appear, the extended clouds very quickly erase spectral features due to being optically thick to nearly the top of the atmosphere.  This also results in the extended cloud models having larger transit depths than the clear or compact cloud models.  For the extended cloud models at $T_{irr} \leq 2500$ K, both ingress and egress spectra are very flat, and the difference in depth between the two sides of transit is due to the larger scale height of the hotter evening limb.  For yet hotter planets ($T_{irr} = 3000 - 3500$ K), the evening limb remains partially cloud-free, resulting in more prominent spectral features in the egress spectrum and more muted features in the ingress spectrum.

The (likely more physically motivated) compact cloud models have somewhat different behavior.  In this case, since the clouds do not automatically extend to the top of the atmosphere, spectral features due to gas-phase molecules remain apparent in all of the models we produce.  However, the strength of these features can be muted relative to the clear-atmosphere models due to the presence of the clouds.  Cloud tops that extend closer to the upper boundary of the model domain will generally have more of a muting effect on the transmission spectra.  This is the case for the middle range of irradiation temperatures explored by our GCM grid.  For the hottest models, no clouds are present.  For the coldest models, the clouds condense deeper down in the atmosphere, which means that even the cloud tops often lie beneath the photosphere set by gaseous absorption.  As a result, we see a maximum impact of clouds occur around $T_{irr} = 2500$ K in the ingress/egress spectra  for the compact cloud realizations. The spectra at both the highest and lowest temperatures in our grid revert to mostly resembling the cloud-free case. 

For both the extended and compact cloud models, the ``nucleation-limited'' treatment of removing some cloud species for their high nucleation barriers (see Section~\ref{sec:method}) has limited impact.  Generally, the nucleation-limited models have slightly smaller transit depths and less muting of spectral features because some of the cloud species are not included in these calculations.  The impact on the modeled transmission spectra is minimal though, with the largest effects seen in the moderately irradiated models.  We remind the reader that the compact nucleation-limited clouds are likely to be the most physically realistic, of the cloud models that we consider in this work based on previous observational and theoretical studies \citep[e.g.][and references therein]{powell2019transit, gao2020aerosol, roman2021clouds}.

\begin{figure*}
    \centering
    \includegraphics[scale=0.35]{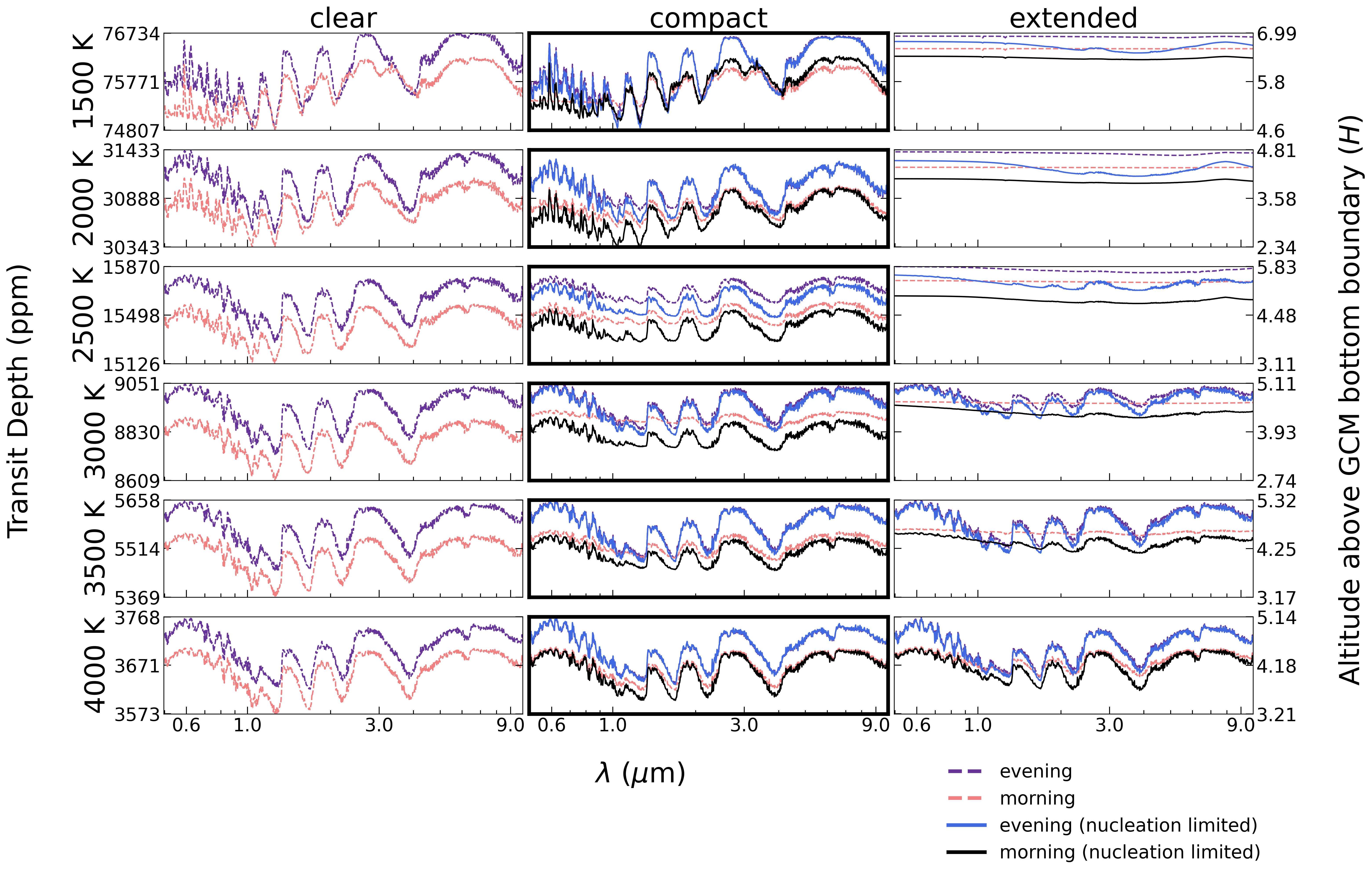}
    \caption{Same as Figure~\ref{fig:transmission grid}, but for morning/evening limb spectra, rather than ingress/egress spectra.}
    \label{fig:transmission grid limb only}
\end{figure*}

\begin{figure*}
    \centering
    \includegraphics[scale=0.35]{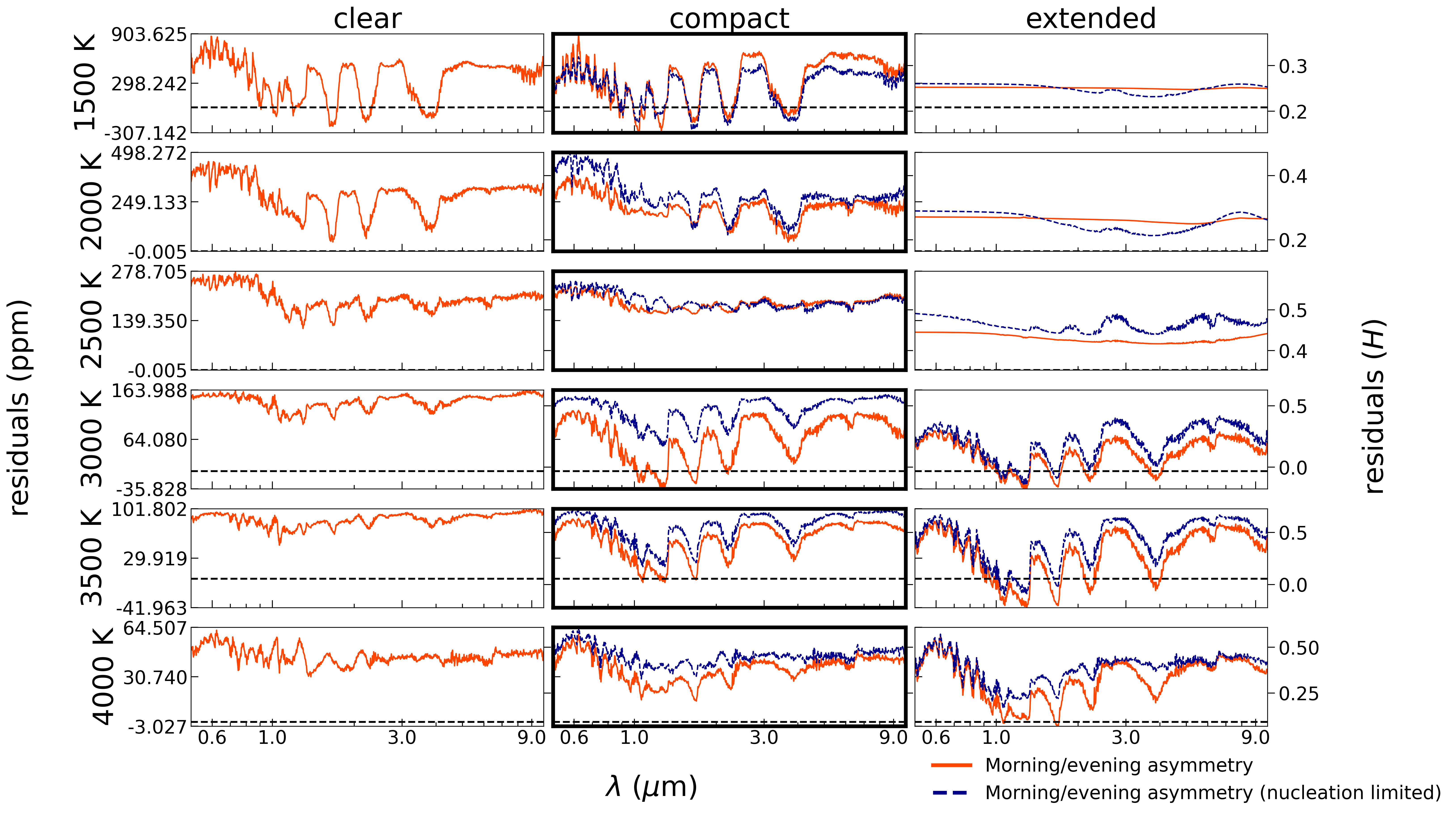}
    \caption{Same as Figure~\ref{fig:residuals grid}, but showing the difference in transit depth between evening and morning limbs, rather than egress and ingress.}
    \label{fig:residuals grid limb only}
\end{figure*}

\begin{figure*}
    \centering
    \includegraphics[scale=0.35]{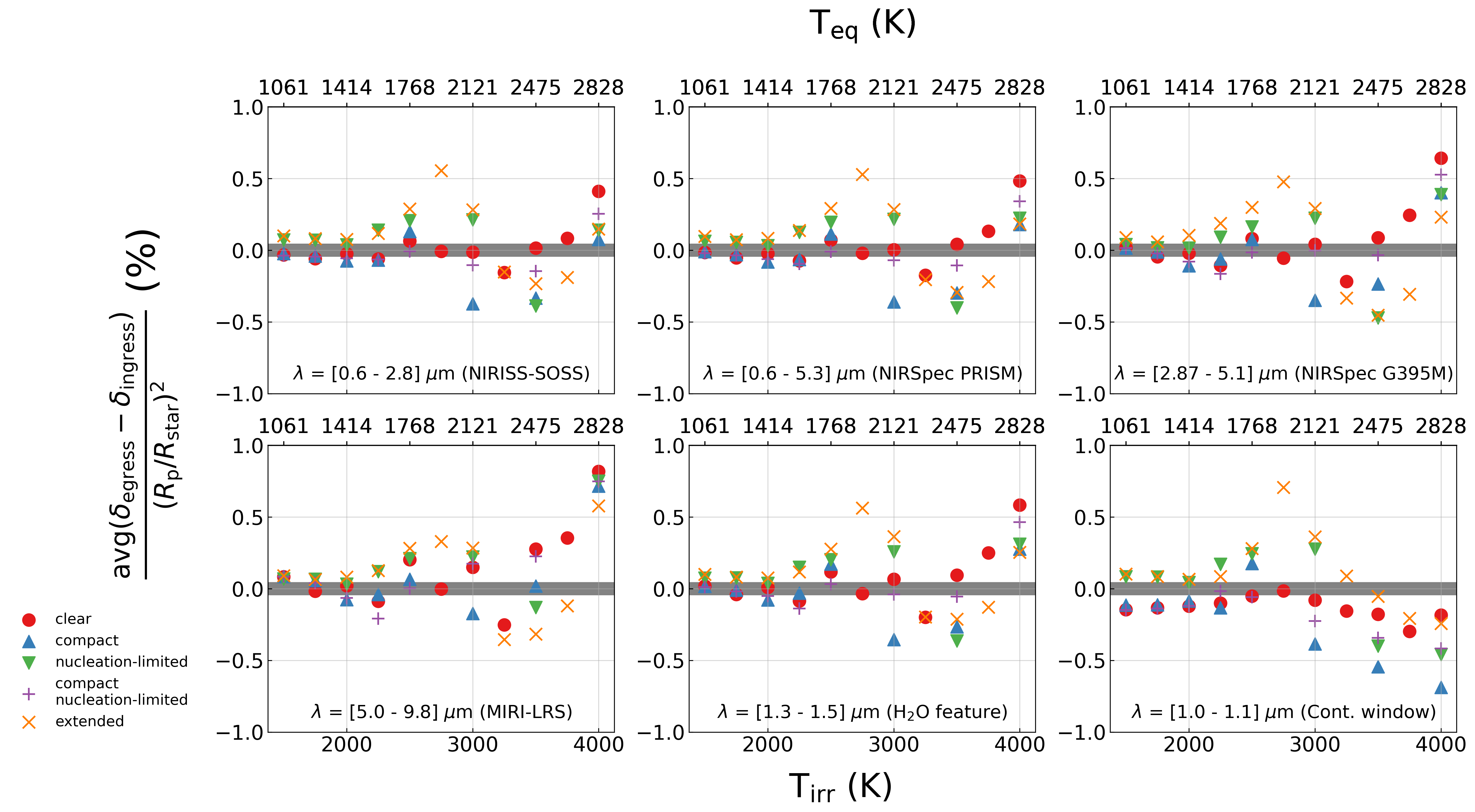}
    \caption{A metric for limb asymmetry strength, averaged across different wavelength ranges accessible to JWST.  The metric is defined as the average difference in transit depth ($\delta$) between egress and ingress, divided by $(\frac{R_{\mathrm{p}}}{R_{\mathrm{star}}})^{2}$, given in percent.  The metric is calculated for models with each cloud treatment, as indicated by the symbols displayed in the plot legend, as a function of planetary irradiation and equilibrium temperature.  Four of the six panels show the metric calculated over the wavelength range of specific JWST instrument modes, as indicated.  The bottom-middle and bottom-right panels show the metric calculated across the wavelength range of a strong H$_2$O feature and a continuum spectral window, respectively. The gray bar indicates the precision on limb asymmetry achieved over the NIRSpec G395H waveband for WASP-39b \cite{espinoza2024inhomogeneous}.}
    \label{fig:metric grid}
\end{figure*}

\subsection{Morning/evening limb transmission spectra}\label{sec:limb only}

In addition to extracting ingress and egress spectra, a second observational technique for measuring limb asymmetry is to attempt to recover separate morning and evening limb spectra using light curve inversion procedures \citep[e.g.][]{powell2019transit, jones20catwoman, espinoza21catwoman}.  These techniques often use simplifications such as assuming that the planet does not rotate during transit, which make the inversion procedure a tractable problem but can also limit their ability to accurately recover the spectra from each hemisphere.  Here we produce modeled morning and evening limb spectra from our GCM grid, in order to compare against the ingress/egress spectra discussed above, and also to diagnose the relative merits of each observational approach.  We generate our morning/evening limb spectra by artificially setting the opacity on one limb of the planet to zero,\footnote{We do, however, preserve the size of the planetary core.} and then performing our ray-striking radiative transfer at an orbital phase of 0$^\circ$ (i.e., center-of-transit).  In Figure~\ref{fig:transmission grid limb only}, we plot the 3D morning/evening spectra in an analogous manner to Figure \ref{fig:transmission grid} for the ingress/egress spectra.  Figure \ref{fig:residuals grid limb only} shows the residuals from subtracting the evening and morning limbs.  

A key difference between the morning/evening vs.\ ingress/egress spectra is that the former typically produces absolute transit depths that are a factor of $\sim$3 larger than the latter.  This occurs because, for our morning/evening spectra, we turn off the opacity of each limb while keeping the \textit{full} planetary core the same size (as opposed to halving it)  to preserve the overall shape of the transit light curve. Otherwise, the spectral shape (vs.\ wavelength) is comparable with each technique, as seen by comparing Figures~\ref{fig:transmission grid} and \ref{fig:transmission grid limb only}.  The \textit{difference} in transit depth between the morning and evening limb spectra is more prominent than between ingress and egress spectra, with the evening limb nearly uniformly producing a larger transit depth than the morning limb. This comes about because the evening limb is the hotter one, and there is no averaging over orbital phase or mixing with the spectrum of the opposing limb when generating morning/evening limb spectra.  As with the ingress/egress spectra, the morning/evening spectra produce transit depth \textit{differences} of up to several hundred ppm (though the morning/evening spectra produce greater differences), making the signal realistically observable with JWST.  

\begin{figure*}
    \centering
    \includegraphics[scale=0.35]{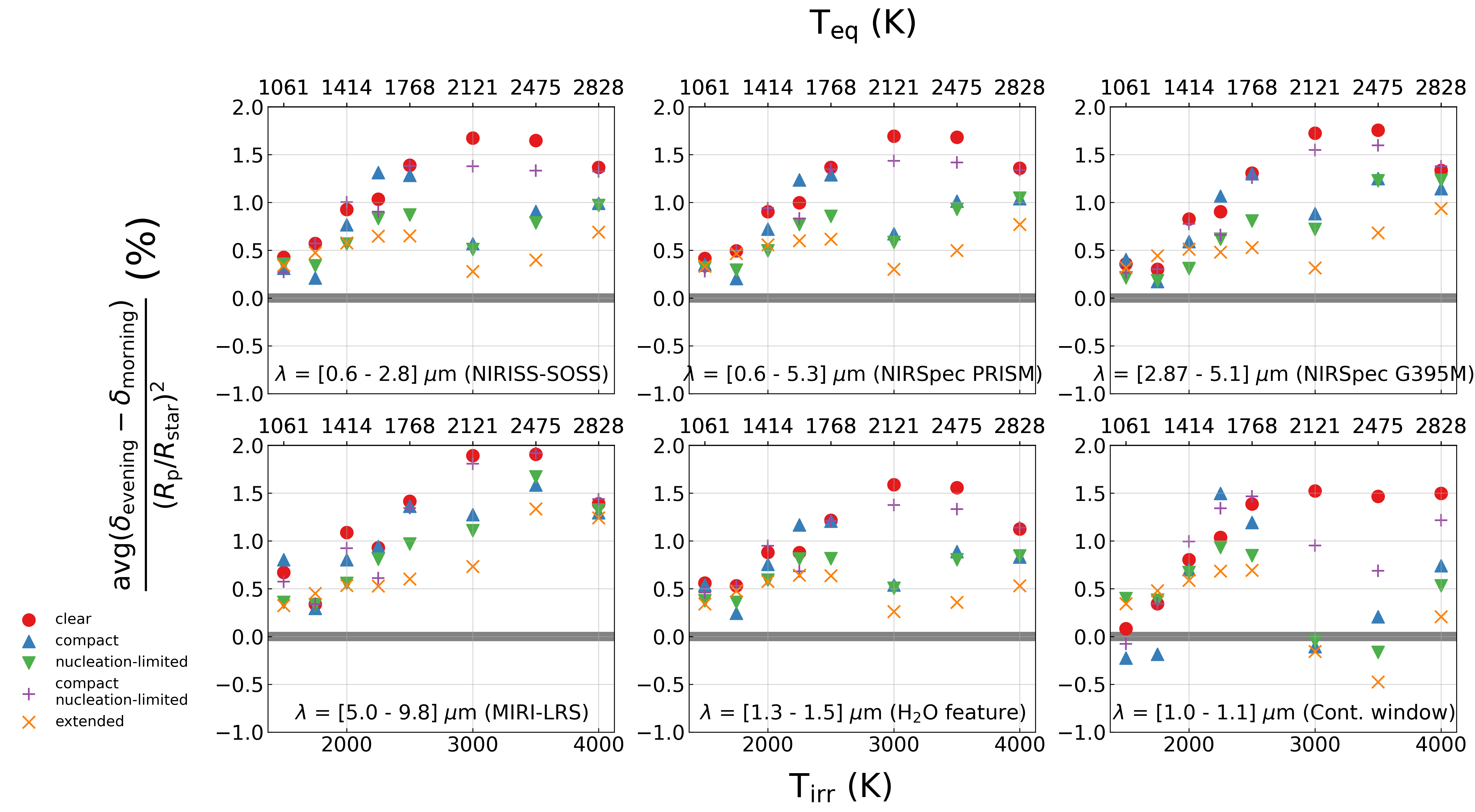}
    \caption{Same as Figure~\ref{fig:metric grid}, but for the difference between evening and morning limb spectra, as opposed to ingress/egress spectra.}
    \label{fig:metric grid limb only}
\end{figure*}

\section{Discussion} \label{sec:discussion}

\subsection{Which wavelength ranges are best for probing limb asymmetry?}

Figures~\ref{fig:residuals grid} and \ref{fig:residuals grid limb only} demonstrate that limb asymmetries should manifest at a level that is observable with JWST (well in excess of 50ppm).  This then raises the question of which JWST observing modes or which wavelength ranges will be most sensitive to diagnosing limb asymmetries.  In Figures~\ref{fig:metric grid} and \ref{fig:metric grid limb only}, we address this question by calculating the average ingress/egress difference and morning/evening difference, respectively, over various characteristic wavelength ranges.  In both cases, we calculate the average difference, divided by the full transit depth ($R_p^2/R_*^2$), to identify the normalized average limb asymmetry.  

First examining Figure~\ref{fig:metric grid} (as well as Figures~\ref{fig:transmission grid} and \ref{fig:residuals grid}), we find that ingress/egress spectra typically produce the largest limb asymmetry signals for planets with higher levels of irradiation.  The sharp uptick in limb asymmetry at high irradiation temperatures arises primarily from the hottest planets rotating through the largest range of orbital phase during transit.  Averaging over each of the JWST instrument modes considered (NIRISS-SOSS, NIRSpec PRISM, NIRSpec G395, and MIRI-LRS), the trend in limb asymmetry vs.\ irradiation temperature is similar. 
 MIRI-LRS produces the largest limb asymmetries on average, approaching a maximum of 1\% of the transit depth (i.e., a bandpass-averaged limb asymmetry of 100 ppm for a hot Jupiter with a 1\% total transit depth). The differences between the various clear and cloudy models is most apparent at irradiation temperatures between 2000 and 3000 K, peaking at 2750 K.  Averaging over smaller wavelength ranges focused specifically on the 1.4 $\mu$m water feature and an opacity window at $1.0-1.1$ $\mu$m, we find a similar overall trend to limb asymmetry vs.\ irradiation temperature, but the differences between the various cloudy and clear models are more pronounced at low irradiation temperatures in the continuum window.  Additionally, the continuum window does not produce the sharp uptick in limb asymmetry at high temperatures that is seen over the other wavelength ranges.

 In Figure~\ref{fig:metric grid limb only} (as well as Figures~\ref{fig:transmission grid limb only} and \ref{fig:residuals grid limb only}), we find a different trend with irradiation temperature.  For the morning/evening limb spectra, the limb asymmetries peak at irradiation temperatures of $3000 - 3500$~K, with a slight reduction at higher temperatures.  The turnover at high levels of irradiation is because the hottest planets have very small hot spot offsets, and the strong day-night temperature gradient in these planets therefore does not translate to a large east-west asymmetry.  
 As with the ingress/egress spectra, the largest limb asymmetry signals occur in the MIRI-LRS bandpass manifesting at a level of 2\% of the total transit depth.  In the morning/evening limb calculations, the cloud signatures are also most apparent at higher irradiation temperatures (between 2500 and 3500 K).  As with the ingress/egress spectra, the $1.0-1.1$ $\mu$m continuum window is most sensitive to the differences among cloud treatments.

Given the similarities in the net limb asymmetry when averaging across the various instrument modes, there is not an obvious choice for which instrument mode should be used to pursue limb asymmetry analyses.  We can conclude that all JWST instruments are useful for measuring limb asymmetries, with MIRI-LRS perhaps being slightly more favorable than the other modes. We can also conclude that focusing on opacity windows is a productive strategy to probe asymmetries in aerosol coverage.  Thicker clouds (i.e., our extended cloud models) produce the largest deviations from clear atmosphere expectations for limb asymmetries. Our limiting-case study motivates the idea that measuring limb asymmetries for a large population of giant planets spanning a range of irradiation temperatures can therefore be useful in identifying the typical thickness of clouds.  

As a caveat, we do note that this paper presents calculations of ``signal,'' but we have as of yet not discussed ``noise''---and signal-to-noise ratios ultimately determine the observability of limb asymmetry with, e.g., JWST. We emphasize that real observations should be planned based on detailed instrumental simulations \citep[e.g.,][]{batalha2017pandexo} that include real sources of uncertainty \citep[e.g.,][]{murphy2024analytic}.

Finally, the trends in limb asymmetries with irradiation temperature for ingress/egress and morning/evening spectra have some notable differences---namely, that the overall asymmetries are smaller. Morning/evening spectra therefore appear to be better suited for measuring limb asymmetries than ingress/egress spectra.  These differences arise because the ingress/egress spectra average over multiple orbital phases as the planet rotates on its axis, and they also include a certain degree of pollution between the signals arising from the morning and evening limbs. Such pollution would be exacerbated in planets with nonzero impact parameters \citep[e.g.,][]{fortney10_3D}. For planets that rotate considerably during transit (i.e., the most highly irradiated planets in our model grid), the planet's rotation tends to exacerbate ingress/egress limb asymmetries. If instead the morning/evening limb spectra can be reliably extracted and corrected for the effects of rotation, we predict a peak temperature of limb asymmetries of $T_{irr}\approx2750$~K across most wavelength ranges.

\subsection{Which physical effects dominate limb asymmetry?}

There are several physical effects that one might expect to impact limb asymmetry measurements for warm-to-ultrahot Jupiters.  Differences in temperature, aerosols, or atmospheric composition between the eastern and western limbs of a planet all have the potential to produce observable differences between the two hemispheres. 
As seen in Section~\ref{sec:results} and in Figures~\ref{fig:transmission grid} -- \ref{fig:metric grid limb only}, we find that temperature differences are the dominant cause of limb asymmetry over the entire domain of our model grid. The larger scale height of the hotter gas on the planet's evening (eastern) terminator leads to larger spectral features in the evening limb or egress spectrum.  The temperature effect alone though should not lead to considerable \textit{wavelength-dependent} limb asymmetries.  All else being equal, for a temperature-independent opacity, if only the gas temperature differs between morning and evening limbs, the limb asymmetry should manifest as an offset between the transit depth obtained on the eastern and western sides of the planet.  This can be seen, for example, in the lower left panels of Figure~\ref{fig:residuals grid limb only}, where the primary east-west asymmetry in the clear atmosphere models for the hottest planets results from temperature only (as opposed to clouds or abundance differences).  Wavelength-dependent limb asymmetries must come about from further differences to the opacity function between the two sides of the planet.

As discussed in Section~\ref{sec:results}, clouds become increasingly responsible for limb asymmetries for the cooler models in our grid.  If the compact clouds are in fact a more realistic depiction of aerosols in hot Jupiter atmospheres, then the cloud effects become most apparent at an irradiation temperature of $\sim2500 - 3000$ K, where the morning limb is more clouded than the evening limb.  Clouds impart a wavelength dependence to the limb asymmetry signature by erasing spectral features on the more clouded limb.  For planets that have equally clouded morning and evening limbs, like our coldest extended-cloud models, the limb asymmetry reverts to being gray in nature, and the magnitude of the limb asymmetry results primarily from temperature differences (see e.g., the upper right panels of Figure~\ref{fig:residuals grid limb only}).  

A third effect that can cause limb asymmetries is east-west gradients in chemical abundances. Our post-processing calculations enforce local thermochemical equilibrium, meaning that temperature differences in the atmosphere can induce gradients in the equilibrium chemical abundances.  Rather, the temperature-dependence of the equilibrium chemical abundances is baked into our models. 
An example of chemical abundance patterns driving limb asymmetries can be seen in the upper left panels of Figures~\ref{fig:transmission grid limb only} and \ref{fig:residuals grid limb only}.  For these coldest clear-atmosphere models, the cooler morning limb dips below temperatures at which methane should become the dominant carbon-bearing molecule in chemical equilibrium.  The resulting CH$_4$ features superimposed over the dominant H$_2$O features across the near-to-mid IR significantly alter the shape of the morning limb spectrum.  This manifests as a strong wavelength dependence in the limb asymmetry signal. Disequilibrium chemical processes \citep[e.g., chemical transport by bulk motion, quenching, or photochemistry;][]{cooper06, agundez2014pseudo,baeyens2022grid,tsai24, zamyatina24} may further enhance or suppress east/west abundance gradients.  Limb asymmetries arising from abundance gradients should manifest as strong wavelength-dependent differences in ingress/egress or morning/evening transmission spectra, corresponding to the wavelength ranges of the specific absorbers that have non-constant abundances across the terminator.  

\subsection{Comparison to previous studies}\label{sec:previous studies}
To further highlight the importance of our models, we draw connections between our models to other works and find applications to previous observations. In \cite{feinstein2023early}, WASP-39b's transmission spectrum was measured with JWST's NIRISS instrument from 0.6 - 2.8 $\mu$m.  Model fits to the data imply partial cloud coverage on the terminator at a level of $50-70$\%.  However, this was a traditional transmission spectroscopy analysis in which phase-dependent information was not extracted, so the location of the clouds (and whether they are localized vs.\ dispersed) cannot be recovered.  Still, the hints of non-uniform cloud coverage imply the likely presence of a limb asymmetry in aerosol coverage that is worth investigating further.

To robustly extract asymmetries from transit light curves, new parameterization techniques are being developed, accompanied by publicly available software such as \texttt{catwoman} \citep{jones20catwoman, espinoza21catwoman} and  \texttt{harmonica} \citep{grant2023tran+}. As an initial proof of concept using data from JWST, \cite{espinoza2024inhomogeneous} applied the \texttt{catwoman} package to separately extract WASP-39b's morning and evening transmission spectra in the near-infrared. They report a significant difference between the evening and morning transit depths, as well as a morning-limb spectrum that is consistent with weaker spectral features, as expected from a colder or cloudier morning terminator. Similarly, \cite{murphy24}  applied \texttt{catwoman} to observations of the warm Jupiter WASP-107b and recovered a larger evening limb and a smaller morning limb. While WASP-107b lies outside the range of our temperature grid, their results agree nicely with ours in that the terminator asymmetry appears to be driven by temperature differences, cloud opacity, and molecular abundances in the upper atmosphere.  Many more hot and warm Jupiters have been or will soon be observed in transit with JWST, and each of these targets in principle presents an opportunity to perform a similar limb asymmetry analysis. 

In addition to applications toward observations, we also look to compare our 3D transmission spectra to other 3D and 1D results. In \cite{caldas2019effects}, the authors showed that the regions probed by transmission extend significantly toward the day and night sides of the planet, due to the oblique path taken by stellar rays through the planetary atmosphere. They produced a 3D radiative transfer model and applied it to a simulation of GJ 1214b to produce synthetic JWST data and showed that comparing against equivalent 1D models of the planet resulted in an error greater than the expected noise. As a related test, we also produce 1D transmission spectra (Figure~\ref{fig:3D vs 1D}) to compare against our 3D ingress/egress and morning/evening spectra. Our 1D spectra are generated by assuming a single TP profile over the entire morning or evening limb, which is equivalent to the equatorial TP profile at a longitude of 90$^\circ$ (evening terminator) or 270$^\circ$ (morning terminator). We find that the 1D models tend to over-predict the expected degree of limb asymmetry by nearly a factor of 2, as can be seen by comparing across the bottom panels of Figure~\ref{fig:3D vs 1D}. The averaging across a broader range of gas temperatures and cloud properties that occurs in the fully 3D calculation tends to mute the degree of limb asymmetry by approximately a factor of two in our models. This serves as a warning that using 1-D parameterizations to predict or retrieve on limb asymmetry measurements may misrepresent the actual properties of the underlying atmosphere. Indeed, \citet{wardenier2022all} find that it is more accurate to average over the TP profiles given an opening angle in \textit{longitude} than to take just the equatorial profiles at 90$^\circ$ and 270$^\circ$.  By similar reasoning, \textit{latitudinal} averaging over the terminator may produce more accurate results as well.  In concurrence with works such as \citet{caldas2019effects}, \citet{nixon2022aura}, \citet{macdonald2022trident}, and \citet{pluriel2023hot}, our results provide further evidence that future modeling efforts may necessitate accounting for 3D effects as observations become more precise. 

\section{Conclusions} \label{sec:conclusion}
We have presented results of 3D ray-striking radiative transfer calculations produced from a grid of hot Jupiter GCMs with radiatively active clouds. These calculations allow us to produce ingress/egress and morning/evening transmission spectra, which we use to simulate JWST data products aimed at extracting east-west limb asymmetries.
Our models are calculated with varying prescriptions for cloud formation, which we use to assess the impact cloud opacity has on the limb asymmetry present in a planet's transmission spectrum. 

We summarize the key results of our modeling study, as follows:
\begin{itemize}
  \item Temperature contrasts, which drive a scale height difference between the morning and evening limbs, are the leading-order effect when considering the strength of asymmetry. All else being equal, a temperature difference alone will not impart a strong wavelength dependence to the limb asymmetry signal.  Wavelength-dependent limb asymmetries can therefore be seen as indicative of other processes such as aerosols or abundance differences playing a role.
  \item We qualitatively predict stronger limb asymmetries to occur for more highly irradiated planets.  This is due to the strong day-night temperature gradient combined with an eastward jet, which drives the east/west asymmetry.  Planetary rotation enhances limb asymmetry in ingress/egress spectra.  For pure morning/evening-limb spectra, the limb asymmetry is expected to peak around $T_{irr} \approx 3000-3500$~K --- hotter planets do not have sufficient eastward heat transport to drive such large east/west temperature differences (e.g., due to high temperatures producing short radiative timescales). These differences are observable in transmission spectra.
  \item Cloud opacity at the lowest and highest irradiation temperatures in our grid does not contribute strongly to the degree of asymmetry. This is because the cloud opacity either completely dominates the planets' optical depth on both limbs, or very few clouds form in the observable atmosphere to impact the spectra in any significant way. The most significant impact of clouds occurs at irradiation temperatures of $\sim2500 - 3000$ K, where the morning limb is primarily clouded, but the hotter evening limb retains clear-atmosphere conditions. 
  \item Key differences exist between measuring limb asymmetries by ingress/egress vs.\ morning/evening limb spectra.  Morning/evening spectra in principle allow for the east and west portions of the terminator to be fully disentangled from one another, which also generally results in a larger predicted limb asymmetry signal.  However, techniques to directly extract morning/evening spectra rely on simplifying assumptions such as a non-rotating planet, which can limit their accuracy in the case in which the planet rotates considerably during transit.  In contrast, ingress/egress spectra are a more directly measurable quantity.  They simultaneously encode information about planetary rotation, and they include some spectral information about the non-dominant limb.  These complicating factors potentially allow for additional information (e.g., about changes with orbital phase) to be extracted, but this requires detailed modeling to accurately infer.  
  \item All JWST observing modes for transiting exoplanet studies have the potential to reveal the signatures of limb asymmetries.  In our modeling, MIRI LRS produced slightly larger limb asymmetry signals on average, but the advantage over other modes is marginal.  In both ingress/egress and morning/evening spectra, we find that measurements made at wavelengths corresponding to opacity windows have the strongest power to distinguish among cloudy vs.\ non-clouded scenarios.  
\end{itemize}

Limb asymmetry analyses now join a growing number of techniques aimed at extracting 3D information from exoplanet spectra.  Phase curves \citep{knutson07}, eclipse mapping \citep{rauscher07}, high-resolution spectroscopy \citep[e.g.][]{ehrenreich10}, and 3D transmission retrieval techniques \citep[e.g.][]{nixon2022aura, macdonald2022trident,falco2024signature} have all been described in the literature at length and have been applied to existing data sets. While prior attempts to measure limb asymmetries with the Hubble Space Telescope (HST) did not return any significant signals \citep[e.g.][]{lothringer22}, the first JWST limb asymmetry measurements are now being reported \citep[e.g.,][]{espinoza2024inhomogeneous, murphy24}.  As we have shown in our work, the predicted limb asymmetry signals should be readily observable for at least a subset of hot Jupiters, either via ingress/egress or morning/evening limb spectroscopy.  This exciting new technique provides a means to effectively spatially resolve the terminator, which is a dynamically interesting region of a tidally locked planet.  By necessity, the terminator has typically been treated as spatially homogeneous (with only a radially varying component) in most transmission spectroscopy analyses.  Yet hot Jupiter GCMs predict strong east-west gradients in temperature, composition, and cloud cover that can now be directly probed using the types of analyses that we have simulated in this paper.

\section{Acknowledgments}\label{sec:acknowledgments}
This research has made use of NASA's Astrophysics Data System Bibliographic Services. The authors acknowledge the University of Maryland super computing resources (\url{http://hpcc.umd.edu}) made available for conducting the research reported in this paper. The authors thank Néstor Espinoza for his thoughtful comments. We also thank the anonymous referee for their comments, which greatly improved the quality of this manuscript. A portion of this research was carried out at the Jet Propulsion Laboratory, California Institute of Technology, under a contract with the National Aeronautics and Space Administration (80NM0018D0004). 

\bibliography{Cloudy_transmission}

\section{Appendix} \label{sec:appendix}

This appendix contains supplementary figures. The figures herein show the TP profiles and optical depth maps of the 3 alternate cloud treatments in our GCM grid (compact, extended, and nucleation-limited models) as well as a 3D-to-1D comparison of the clear models. We remind the reader that the compact nucleation-limited cloud scenarios highlighted in the main text constitute our favored physical model.  However, the extended, compact, and nucleation limited models are alternate cloud realizations and still provide insight into cloudy hot Jupiters and the effects of cloud opacity on transmission spectra at various irradiation temperatures.

\renewcommand{\thefigure}{A\arabic{figure}}

\setcounter{figure}{0}

\begin{center}

\begin{figure*}[h!]
    \centering
    \includegraphics[scale=0.4]{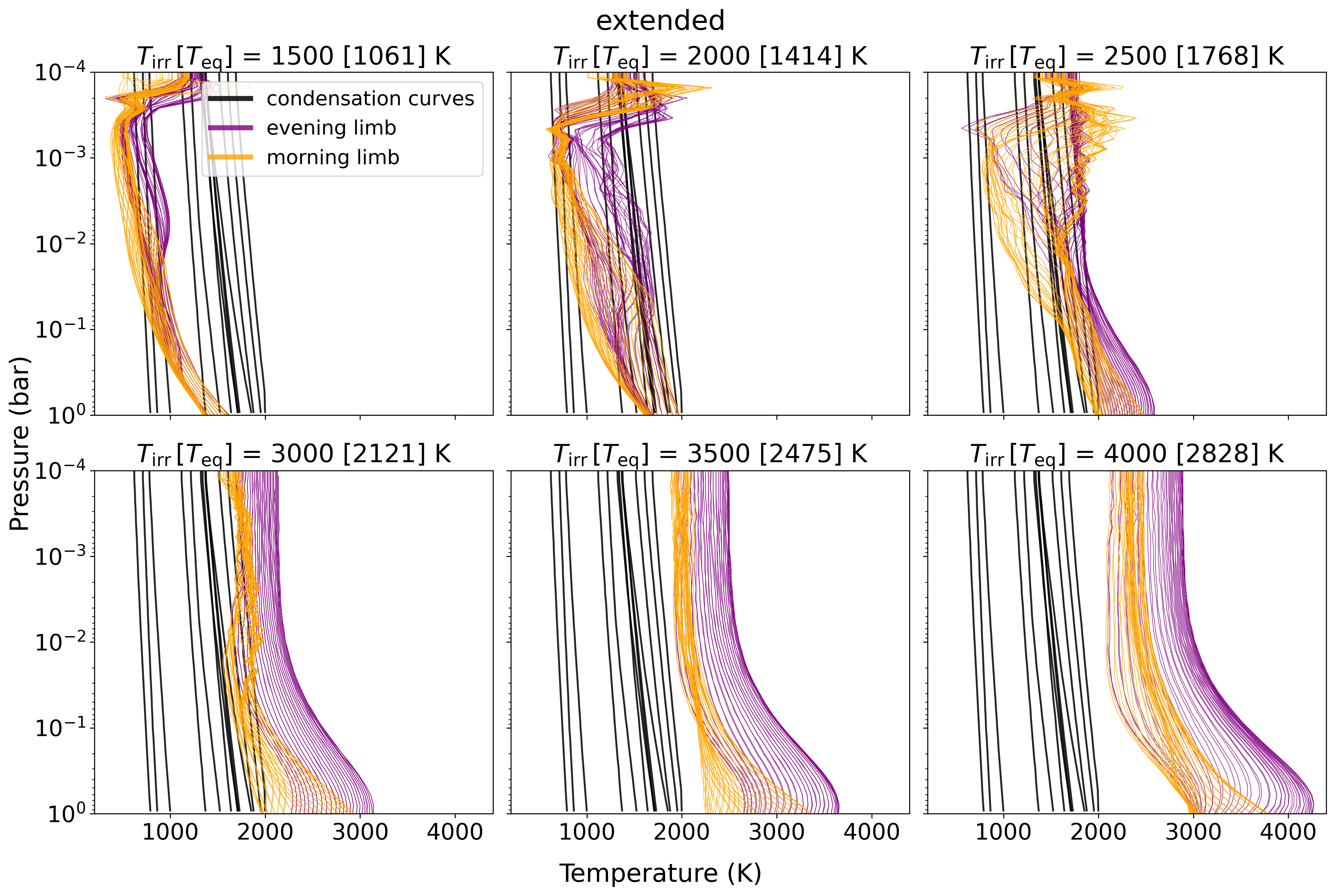}
    \caption{Same as Figure~\ref{fig:clear TP-profile} but for the extended cloud GCMs.}
    \label{fig:extended TP-profile}
\end{figure*}

\begin{figure*}
    \centering
    \includegraphics[scale=0.4]{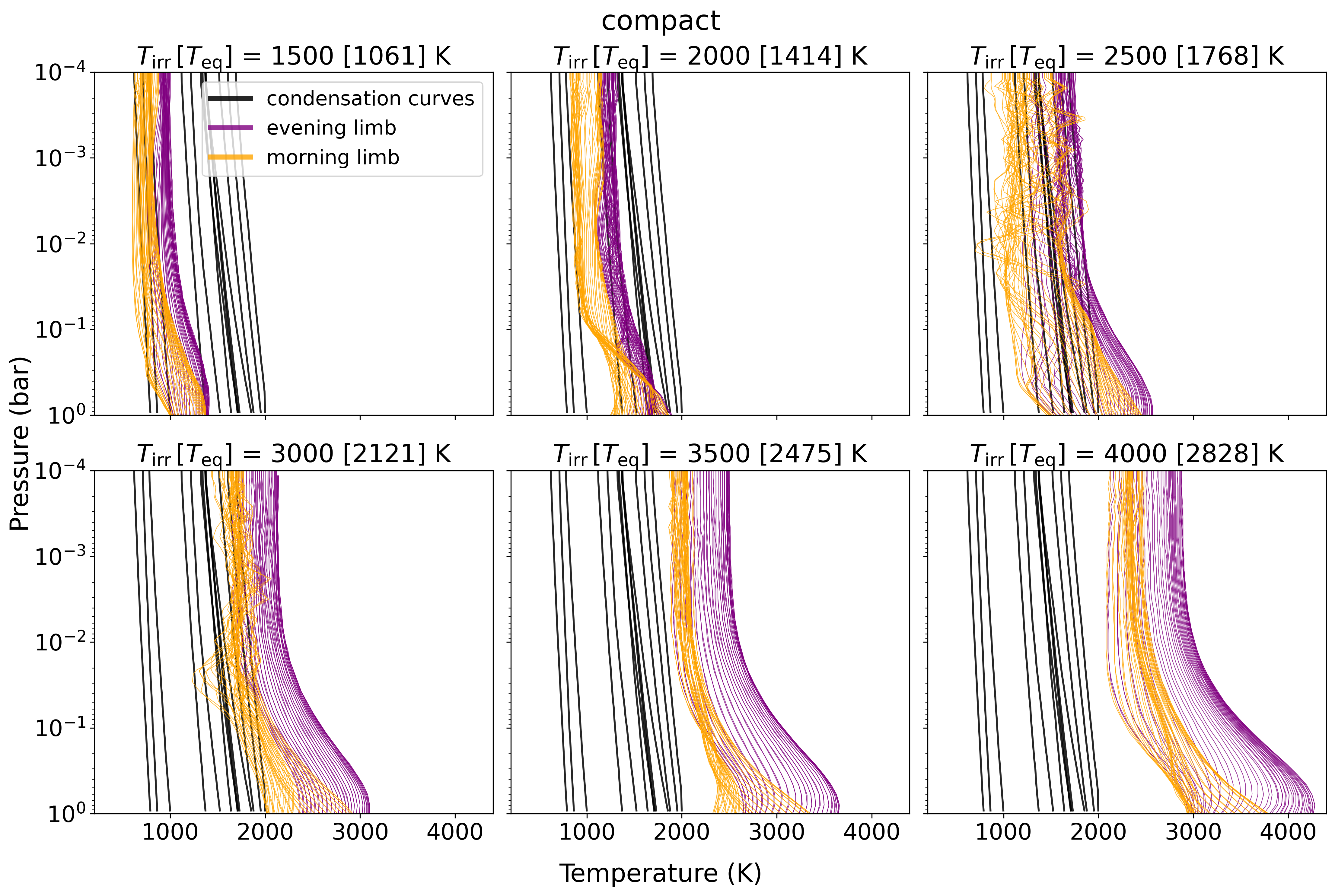}
    \caption{Same as Figure~\ref{fig:clear TP-profile} but for the compact cloud GCMs.}
    \label{fig:compact TP-profile}
\end{figure*}

\begin{figure*}
    \centering
    \includegraphics[scale=0.4]{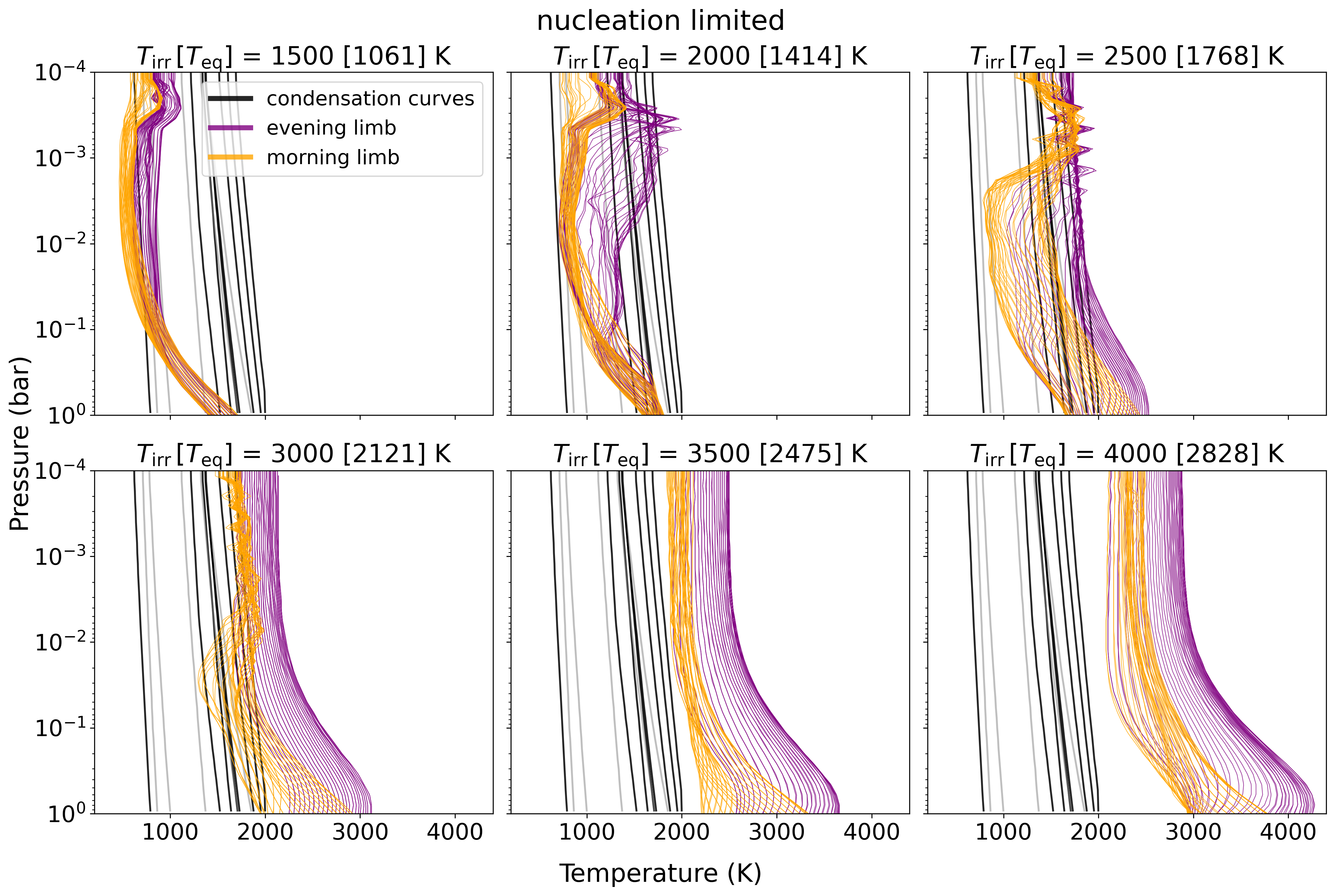}
    \caption{Same as Figure~\ref{fig:clear TP-profile} but for the nucleation-limited (extended cloud) GCMs.}
    \label{fig:nuclim TP-profile}
\end{figure*}

\begin{figure*}
    \centering
    \includegraphics[scale=0.8]{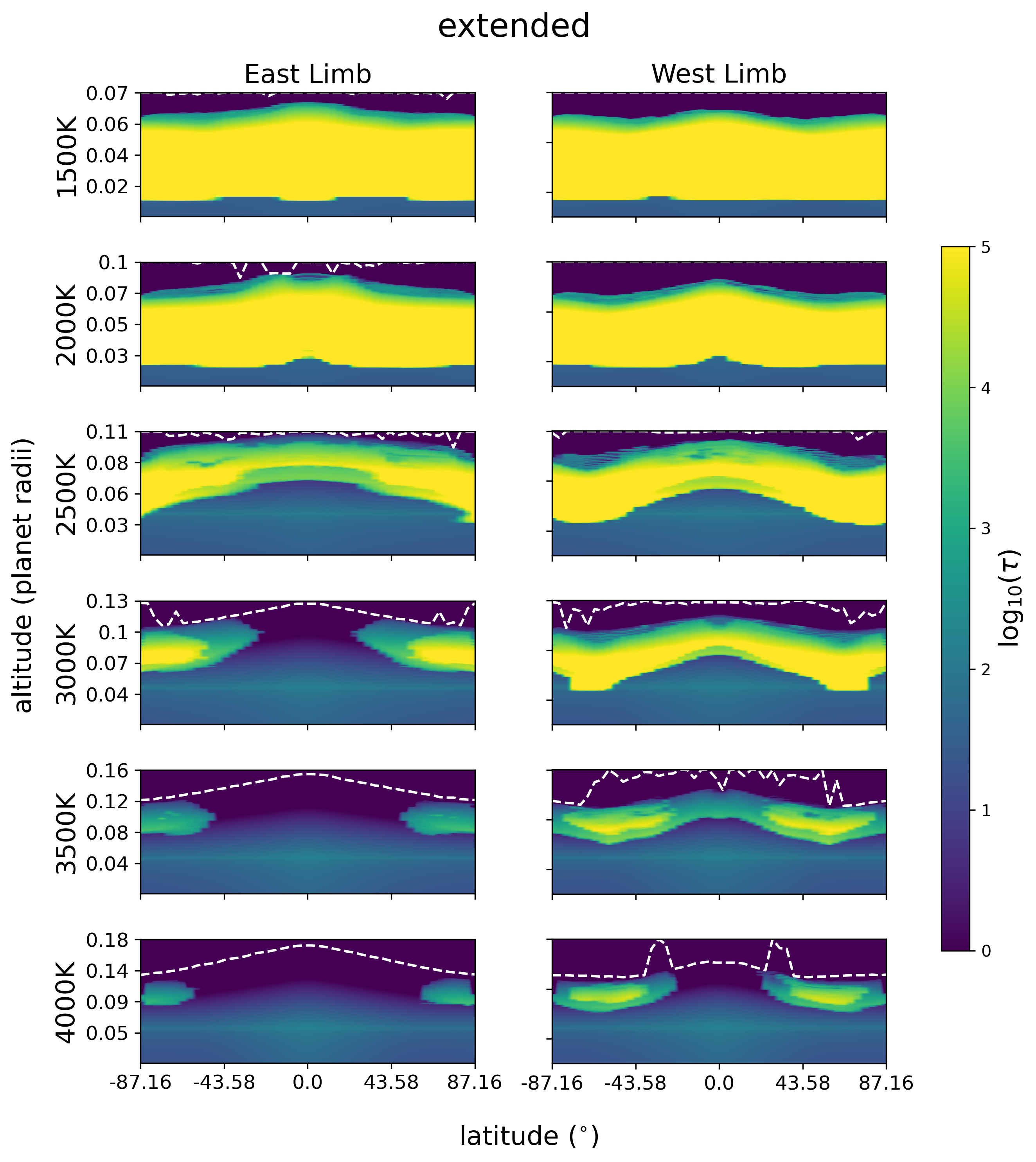}
    \caption{Same as Figure~\ref{fig:ODM cartesian comnuclim}, but for the extended cloud models.}
    \label{fig:ODM cartesian extended}
\end{figure*}

\begin{figure*}
    \centering
    \includegraphics[scale=0.8]{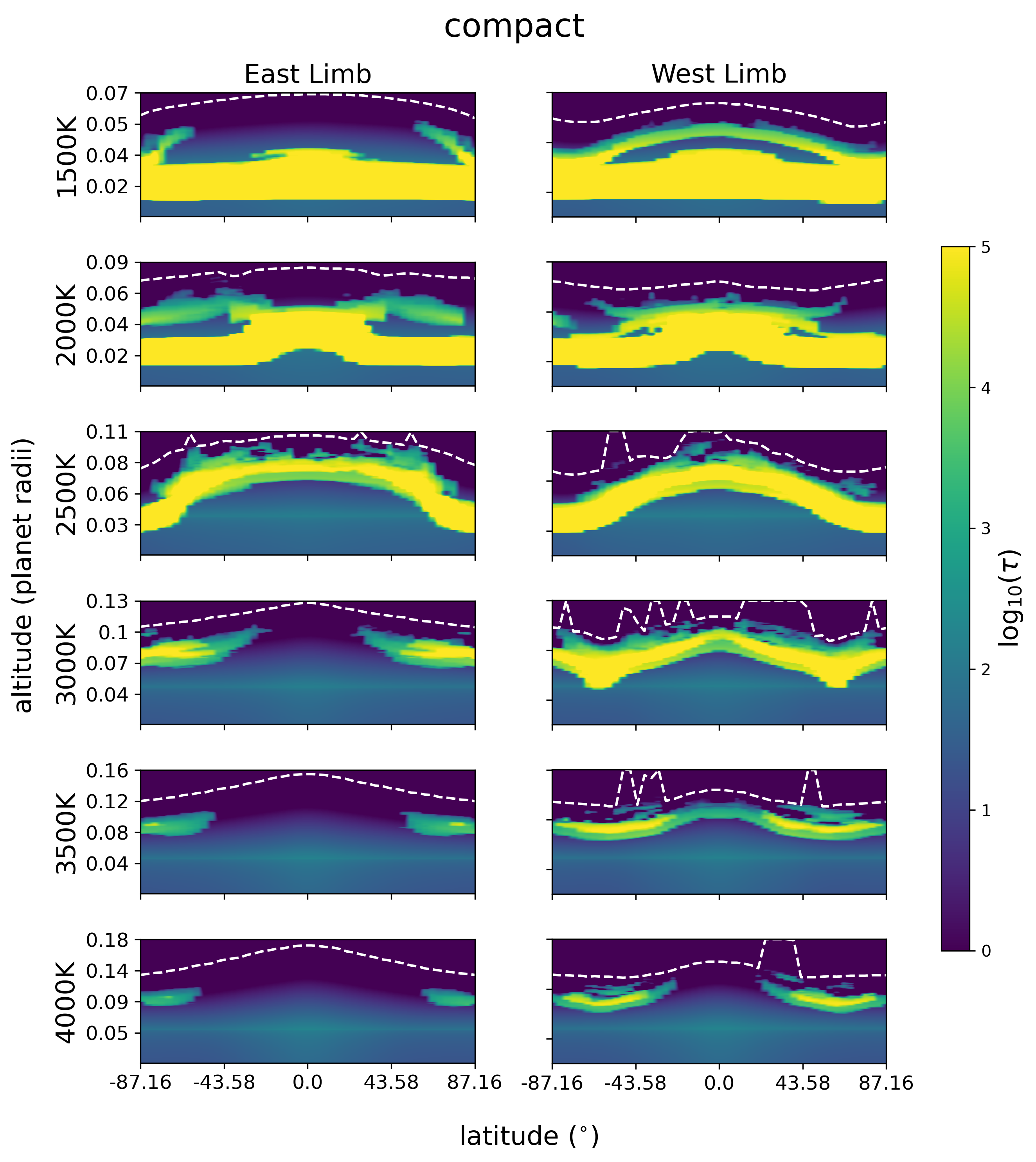}
    \caption{Same as Figure~\ref{fig:ODM cartesian comnuclim}, but for the compact cloud models.}
    \label{fig:ODM cartesian compact}
\end{figure*}

\begin{figure*}
    \centering
    \includegraphics[scale=0.8]{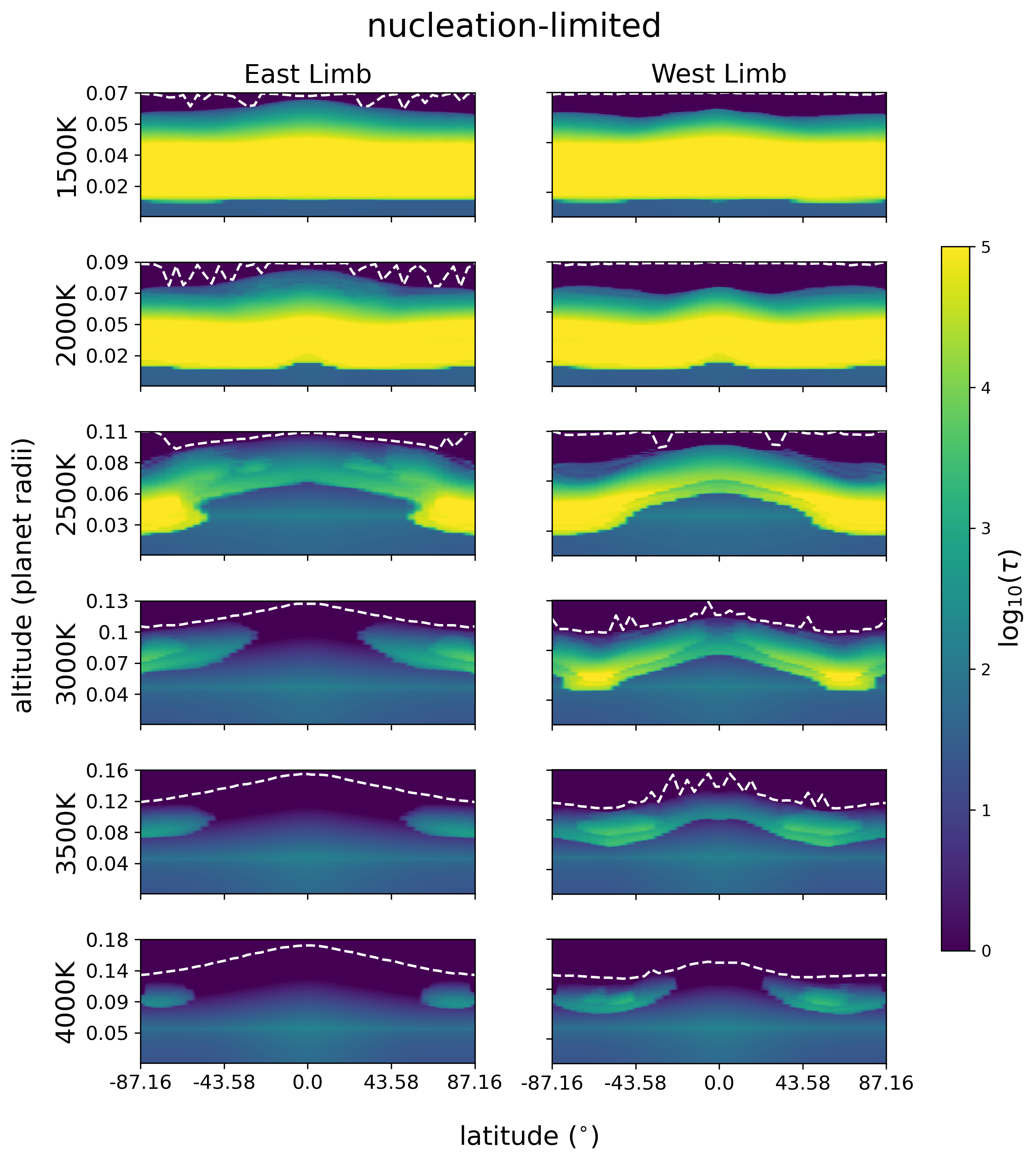}
    \caption{Same as Figure~\ref{fig:ODM cartesian comnuclim}, but for the nucleation-limited (extended) cloud models.}
    \label{fig:ODM cartesian nuclim}
\end{figure*}

\begin{figure*}
    \centering
    \includegraphics[scale=0.58]{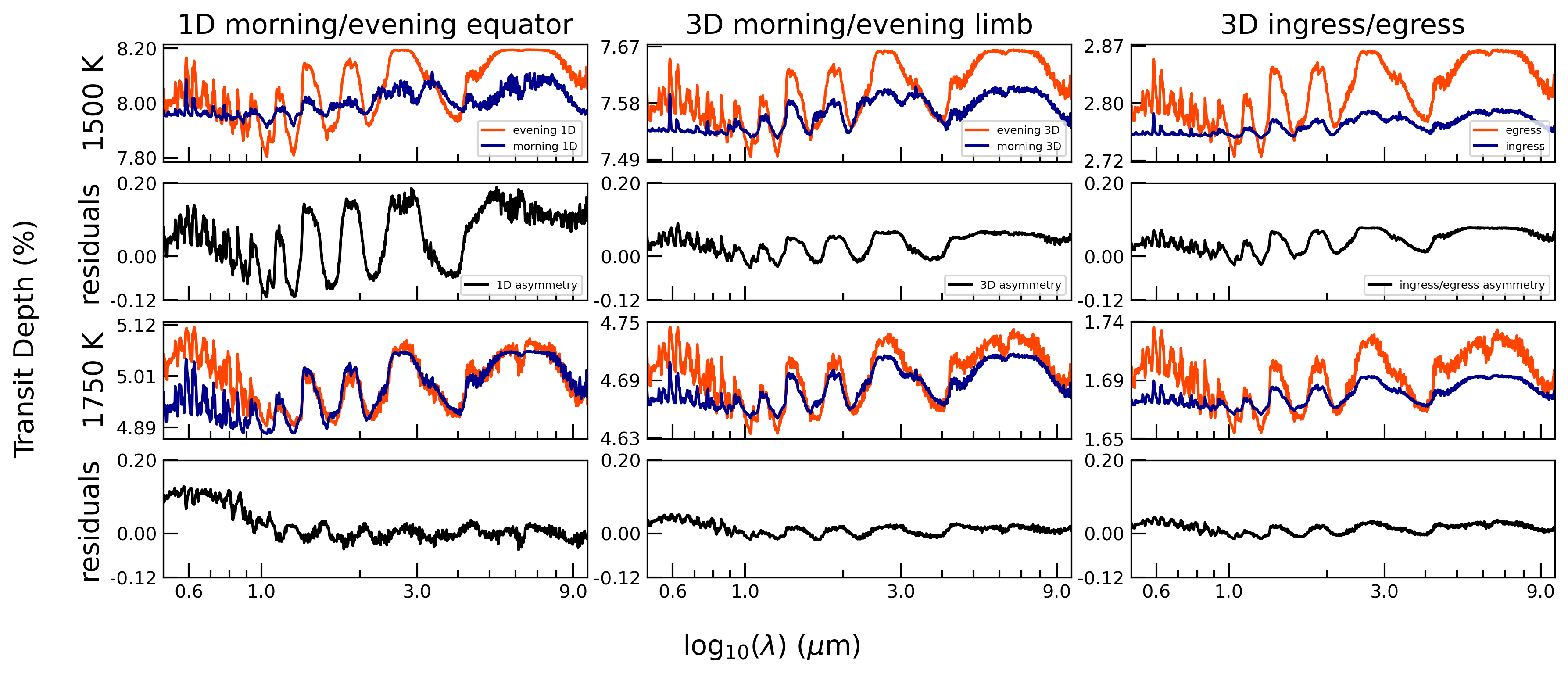}
    \caption{Comparison between results obtained for 1D morning/evening equator spectra, vs.\ 3D morning/evening spectra and 3D ingress/egress spectra. The spectra in this grid are from the compact nucleation-limited cloud GCMs, simulated for $T_{\mathrm{irr}} = 1500\textnormal{K}$ and $T_{\mathrm{irr}} = 1750\textnormal{K}$. The 1-D spectra are obtained by assuming a uniform morning (or evening) terminator, using the equatorial terminator T-P profile.  Residuals are plotted as the difference between morning and evening (or ingress and egress) spectra.  The 1D calculation over-estimates the strength of limb asymmetries by a factor of $\sim$2, relative to the 3D simulations. Note the differing y-axis ranges for each panel.}
    \label{fig:3D vs 1D}
\end{figure*}
\end{center}

\end{document}